\documentclass[conference,svgnames]{vgtc}                     

\onlineid{0}

\vgtccategory{Research}

\vgtcpapertype{please specify}

\title{The State of Reproducibility Stamps for Visualization Research Papers}

\author{%
  \authororcid{Tobias Isenberg}{0000-0001-7953-8644}\thanks{e-mail: given\_name.family\_name@inria.fr}\\ %
        \scriptsize Université Paris-Saclay, CNRS, Inria, LISN, France%
}

\authorfooter{
  \item
  	Tobias Isenberg is with Université Paris-Saclay, CNRS, Inria, LISN, France. E-mail: given\_name.family\_name@inria.fr
}

\abstract{%
	I analyze the evolution of papers certified by the  Graphics Replicability Stamp Initiative (GRSI) to be reproducible, with a specific focus on the subset of publications that address visualization-related topics. With this analysis I show that, while the number of papers is increasing overall and within the visualization field, we still have to improve quite a bit to escape the replication crisis. I base my analysis on the data published by the GRSI as well as publication data for the different venues in visualization and lists of journal papers that have been presented at visualization-focused conferences. I also analyze the differences between the involved journals as well as the percentage of reproducible papers in the different presentation venues. Furthermore, I look at the authors of the publications and, in particular, their affiliation countries to see where most reproducible papers come from. Finally, I discuss potential reasons for the low reproducibility numbers and suggest possible ways to overcome \mbox{these obstacles. This paper is reproducible itself, with source} code and data available from \href{https://github.com/tobiasisenberg/Visualization-Reproducibility}{\texttt{github\discretionary{}{.}{.}com\discretionary{/}{}{/}tobiasisenberg\discretionary{/}{}{/}Visualization\discretionary{}{-}{-}Reproducibility}} as well as a free paper copy and all supplemental materials at \href{https://osf.io/\osfid/}{\texttt{osf.io/\osfid}}.%
}


\teaser{
  \centering
  \includegraphics[width=\linewidth, alt={A view of a city with buildings peeking out of the clouds.}]{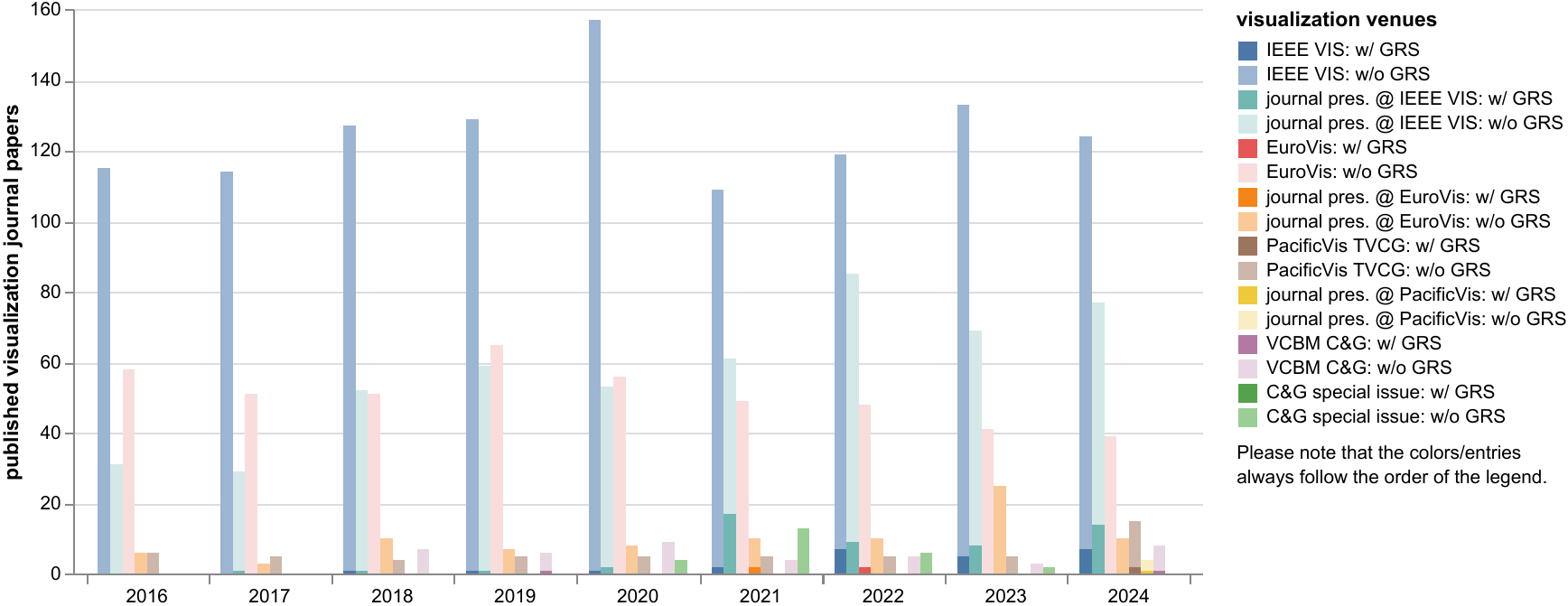}
  \caption{Evolution of publication counts in several visualization venues,\textsuperscript{\ref{foot:vis_presentations},\ref{foot:c-and-g-special}} including their proportion of papers with GRS certification (for a normalized version with percentages see \autoref{fig:vis-grs-presentations-by-year}). We can see that only few papers so far have received a GRS, in all venues.}
  \label{fig:teaser}
}

\graphicspath{{../graphs/}{../graphs 20240603/}{figs/}{figures/}{pictures/}{images/}{./}} 

\usepackage{ccicons}
\usepackage{nicefrac}
\usepackage{mathptmx}                  
\usepackage{pgf}

\newcommand*{\percentageRounded}[2]{%
	\pgfmathparse{#1*100/#2}%
	\pgfmathprintnumber[fixed, precision=1]{\pgfmathresult}%
}

\newcommand{\eg}{e.\,g.}
\newcommand{\ie}{i.\,e.}

\newcommand{\osfid}{mvnbj}

\newcommand{\GrsiCountryPieChartThreshold}{2.5}

\newcommand{\TotalIeeeVisPapersInMMXXI}{109}
\newcommand{\TotalIeeeVisPapersInMMXXII}{119}
\newcommand{\TotalIeeeVisPapersInMMXXIII}{133}
\newcommand{\TotalIeeeVisPapersInMMXXIV}{124}

\newcommand{\TotalIeeeVisTVCGJournalPapersInMMXXI}{50}
\newcommand{\TotalIeeeVisTVCGJournalPapersInMMXXII}{67}
\newcommand{\TotalIeeeVisTVCGJournalPapersInMMXXIII}{60}
\newcommand{\TotalIeeeVisTVCGJournalPapersInMMXXIV}{65}

\newcommand{\GrsiDataCurrentAsOf}{Sep.~12, 2024}

\newcommand{\GrsiTotalAuthors}{1283}
\newcommand{\GrsiTotalPapers}{360}
\newcommand{\GrsiTotalVisAuthors}{365}
\newcommand{\GrsiIeeeVisPapersCount}{24}

\newcommand{\GrsiIeeeVisPapersInMMXXI}{2}
\newcommand{\GrsiIeeeVisPapersInMMXXII}{7}
\newcommand{\GrsiIeeeVisPapersInMMXXIII}{5}
\newcommand{\GrsiIeeeVisPapersInMMXXIV}{7}
\newcommand{\GrsiIeeeVisJournalPresentationsCount}{53}

\newcommand{\GrsiIeeeVisTVCGJournalPapersInMMXXI}{17}
\newcommand{\GrsiIeeeVisTVCGJournalPapersInMMXXII}{9}
\newcommand{\GrsiIeeeVisTVCGJournalPapersInMMXXIII}{8}
\newcommand{\GrsiIeeeVisTVCGJournalPapersInMMXXIV}{14}
\newcommand{\GrsiIeeeVisTvcgJournalPresentationsCount}{53}
\newcommand{\GrsiPacificVisTvcgPapersCount}{2}

\newcommand{\GrsiPacificVisJournalPresentationsCount}{1}

\newcommand{\GrsiEuroVisPapersCount}{2}

\newcommand{\GrsiEuroVisJournalPresentationsCount}{2}

\newcommand{\GrsiVcbmCagPapersCount}{2}

\newcommand{\GrsiCagSpecialIssuesPapersCount}{0}

\newcommand{\GrsiVisByKeywordPapersCount}{10}
\newcommand{\GrsiVisManuallyMarkedPapersCount}{9}
\newcommand{\GrsiVisKeywordPlusManualPapersCount}{19}
\newcommand{\GrsiTotalVisPapers}{105}
\newcommand{\GrsiVisKeywordPlusManualPapersPercentage}{18.1}
\newcommand{\GrsiPercentageVisPapers}{29.2}
\newcommand{\GrsiDifferenceInPaperDatabases}{0}
\newcommand{\GrsiVisPapersInIEEETVCGTotal}{90}
\newcommand{\GrsiVisPapersInIEEETVCGPresentation}{80}

\newcommand{\GrsiVisPapersInIEEETVCGPercentagePresentation}{88.9}
\newcommand{\GrsiVisPapersInACMToGTotal}{1}
\newcommand{\GrsiVisPapersInACMToGPresentation}{0}

\newcommand{\GrsiVisPapersInACMToGPercentagePresentation}{0.0}
\newcommand{\GrsiVisPapersInWileyCGFTotal}{6}
\newcommand{\GrsiVisPapersInWileyCGFPresentation}{2}

\newcommand{\GrsiVisPapersInWileyCGFPercentagePresentation}{33.3}
\newcommand{\GrsiVisPapersInElsevierCaGTotal}{8}
\newcommand{\GrsiVisPapersInElsevierCaGPresentation}{4}

\newcommand{\GrsiVisPapersInElsevierCaGPercentagePresentation}{50.0}
\newcommand{\GrsiCountryPieChartOverallNoOneName}{the United States}
\newcommand{\GrsiCountryPieChartOverallNoOnePercentage}{21.1}
\newcommand{\GrsiCountryPieChartOverallNoTwoName}{China}
\newcommand{\GrsiCountryPieChartOverallNoTwoPercentage}{15.3}
\newcommand{\GrsiCountryPieChartOverallNoThreeName}{France}
\newcommand{\GrsiCountryPieChartOverallNoThreePercentage}{15.1}
\newcommand{\GrsiCountryPieChartOverallNoFourName}{Germany}
\newcommand{\GrsiCountryPieChartOverallNoFourPercentage}{11.6}

\newcommand{\GrsiCountryPieChartVisNoOneName}{the United States}
\newcommand{\GrsiCountryPieChartVisNoOnePercentage}{25.2}
\newcommand{\GrsiCountryPieChartVisNoTwoName}{France}
\newcommand{\GrsiCountryPieChartVisNoTwoPercentage}{19.7}
\newcommand{\GrsiCountryPieChartVisNoThreeName}{Germany}
\newcommand{\GrsiCountryPieChartVisNoThreePercentage}{18.1}
\newcommand{\GrsiCountryPieChartVisNoFourName}{China}
\newcommand{\GrsiCountryPieChartVisNoFourPercentage}{11.5}

\newcommand{\GrsiCountryPieChartOverallSeniorNoOneName}{the United States}
\newcommand{\GrsiCountryPieChartOverallSeniorNoOnePercentage}{21.8}
\newcommand{\GrsiCountryPieChartOverallSeniorNoTwoName}{France}
\newcommand{\GrsiCountryPieChartOverallSeniorNoTwoPercentage}{15.7}
\newcommand{\GrsiCountryPieChartOverallSeniorNoThreeName}{China}
\newcommand{\GrsiCountryPieChartOverallSeniorNoThreePercentage}{13.8}
\newcommand{\GrsiCountryPieChartOverallSeniorNoFourName}{Germany}
\newcommand{\GrsiCountryPieChartOverallSeniorNoFourPercentage}{12.1}

\newcommand{\GrsiCountryPieChartVisSeniorNoOneName}{the United States}
\newcommand{\GrsiCountryPieChartVisSeniorNoOnePercentage}{25.2}
\newcommand{\GrsiCountryPieChartVisSeniorNoTwoName}{France}
\newcommand{\GrsiCountryPieChartVisSeniorNoTwoPercentage}{21.4}
\newcommand{\GrsiCountryPieChartVisSeniorNoThreeName}{Germany}
\newcommand{\GrsiCountryPieChartVisSeniorNoThreePercentage}{18.1}
\newcommand{\GrsiCountryPieChartVisSeniorNoFourName}{China}
\newcommand{\GrsiCountryPieChartVisSeniorNoFourPercentage}{11.0}

\begin{document}


\firstsection{Introduction}

\maketitle

``A scientific result is not truly established until it is independently confirmed'' \cite{Boisvert:2016:IR}---yet in the past this mantra has not really been the guiding principle for work within visualization. BELIV as a venue has looked at this question specifically in its 2018 edition \cite{Haroz:2018:OPV,Kosara:2018:SRC,Sukumar:2018:TDU,Valdez:2018:RRR}, and found that there is still much to be improved in our field. Two years later in 2020, Fekete and Freire \cite{Fekete:2020:ERV} observed that ``there are few visualization articles with a [graphics replicability] stamp (6 for TVCG, 4 for CGF), and most of these are about computer graphics.'' They referred to the badges (which, in this paper, I call graphics replicability\footnotemark{} stamps, GRS) awarded by the Graphics Replicability Stamp Initiative (GRSI; \href{https://www.replicabilitystamp.org/}{\texttt{replicabilitystamp\discretionary{}{.}{.}org}}) to published journal\footnotemark{} papers within the larger computer graphics research field for ensuring that some of the presented work can indeed be replicated by independent others. As the GRSI started from within the computer graphics community and had only started to award papers in 2017, only few authors---in particular within the visualization community---up to 2020 really had the time to take advantage of this independent verification process, so the low numbers at the time are no surprise. So now, in 2024, with a lot of past and ongoing discussion within our field on how we can make work more reproducible, I think it is time to take another look and check again, to see if the status quo has improved since then.

\addtocounter{footnote}{-1}
\footnotetext{See the discussion of terminology in \autoref{sec:terminology}.}
\addtocounter{footnote}{1}
\footnotetext{The GRSI started with only certain journals as eligible publication platforms, but in 2024 also added SIGGRAPH and SIGGRAPH Asia conference-only papers as possible venues for papers to receive a GRS. In this paper I still often use ``journal'' to refer to the \emph{publication} venues awarded by the GRSI, in particular since within visualization it is still true as of today. But notice that in the future other non-journal venues may be added.}

Below I describe my work to collect the respective data, from the GRSI website, from the various digital libraries, from conference venues, and from other sources. Based on this data I then set out to analyze not only the GRSI-awarded work overall but also what part of it covers visualization work, how the various publication and presentation venues differ from each other, and how reproducibility is fostered in the different home countries of the researchers.

For full disclosure I note that I am independent\footnote{But my wife currently serves as a reviewer for the GRSI.} of the GRSI, but that I am personally interested in fostering reproducibility and replicability in the field. My collaborators and I did apply for and received several stamps from the GRSI for our work\footnote{\href{https://tobias.isenberg.cc/reproducibility}{\texttt{tobias.isenberg.cc/reproducibility}}} (and also encouraged others to do so)---and I thus also personally contributed to the numbers I report below, in the sense of trying to increase them.

\section{Terminology and scope}
\label{sec:terminology}

Unfortunately, the terminology with respect to being able to check the correctness of scientific results is far from clear \cite{Plesser:2018:RRB}, in particular in the GRSI case. Originally, \emph{reproducibility} was defined to mean that one runs author-provided programs or tools to re-create the same results (e.g., images) using the author-provided input data \cite{Claerbout:1992:EDG}, while \emph{replicability} refers to ``writing and then running new software based on the description of a computational model or method provided in the original publication, and obtaining results that are similar enough to be considered equivalent'' \cite{Rougier​:2017:SCS}. The ACM, however, initially used the exact opposite definition of the terminology: \emph{reproducibility} to mean ``different team, different experimental setup'' and \emph{replicability} to mean ``different team, same experimental setup'' \cite{Boisvert:2016:IR,Plesser:2018:RRB},\footnote{\href{https://www.acm.org/publications/policies/artifact-review-badging}{\texttt{acm.org\discretionary{/}{}{/}publications\discretionary{/}{}{/}policies\discretionary{/}{}{/}artifact\discretionary{}{-}{-}review\discretionary{}{-}{-}badging}}} adding the notion of \emph{repeatability} for a re-creation of the same experimental setup by the original team (also see the discussion of the terminology in the experimental sciences by Plesser \cite{Plesser:2018:RRB}). Luckily, as of August 24, 2020, the ACM aligned their definitions with the original ones, \ie, using \emph{reproducibility} to mean ``different team, same experimental setup'' and \emph{replicability} ``different team, different experimental setup.''\footnote{\href{https://www.acm.org/publications/policies/artifact-review-and-badging-current}{\texttt{acm.org\discretionary{/}{}{/}publications\discretionary{/}{}{/}policies\discretionary{/}{}{/}artifact\discretionary{}{-}{-}review\discretionary{}{-}{-}and\discretionary{}{-}{-}badging\discretionary{}{-}{-}current}}} As the Graphics \emph{Replicability} Stamp Initiative was established in 2016, it likely followed the original definitions of the ACM and thus, despite using the term \emph{replicability}, its focus is to verify independently (\ie, ``different team'') that the author-provided code and data makes it possible to re-create the same results as the authors presented in the paper (\ie, ``same experimental setup''). In this paper, however, to avoid further confusion I follow ACM's updated notion as I discuss the independent verification of the author's code by the GRSI with (usually) the authors' data, using the \emph{reproducibility} terminology---even in the title of this paper.

\section{Related work}
\label{sec:rw}

What motivates this discussion of reproducibility is the realization that way we traditionally conduct and report scientific research is subject to a \emph{replication crisis} \cite{Kosara:2018:SRC,Cockburn:2020:TRC}---work, once published, is rarely questioned or re-checked. Kosara \cite{Kosara:2016:EBS} made this point for the field of visualization by describing it as ``an empire built on sand,'' giving many examples of where established canon should be questioned and re-examined. Yet in our field's review process such replications often are considered to be of too low novelty to be publishable \cite{Quadri:2019:YCP}, such that it requires extensions of pure replications for making the cut as Quadri and Rosen \cite{Quadri:2019:YCP} and even the VIS 2024 overall paper chairs\footnote{\href{https://ieeevis.org/year/2024/blog/vis-2024-OPC-blog-replication}{\texttt{ieeevis\discretionary{}{.}{.}org\discretionary{/}{}{/}year\discretionary{/}{}{/}2024\discretionary{/}{}{/}blog\discretionary{/}{}{/}vis\discretionary{}{-}{-}2024\discretionary{}{-}{-}OPC\discretionary{}{-}{-}blog\discretionary{}{-}{-}replication}}} recommend. Kindlmann\footnote{\href{https://people.cs.uchicago.edu/~glk/talks/pdf/Kindlmann-ScienceInVisualization-VIS-2006-talk.pdf}{\texttt{people\discretionary{}{.}{.}cs\discretionary{}{.}{.}uchicago\discretionary{}{.}{.}edu\discretionary{/}{}{/}\~{}glk\discretionary{/}{}{/}talks\discretionary{/}{}{/}pdf\discretionary{/}{}{/}Kindlmann\discretionary{}{-}{-}ScienceInVisualization\discretionary{}{-}{-}VIS\discretionary{}{-}{-}2006\discretionary{}{-}{-}talk\discretionary{}{.}{.}pdf}}} at the VIS panel ``Is There Science in Visualization?'' already in 2006 argued the same point, and pointed to code availability for published papers as a pre-requisite.

In the past, this code availability that makes reproducibility and replicability possible has not been a core focus of authors. For the computer graphics field, Bonneel et al. \cite{Bonneel:2020:CRC} analyzed the reproducibility\footnote{Also here, different from the paper, I use ACM's updated terminology.} of results published in ToG-level papers from SIGGRAPH 2014, 2016, and 2018, not only manually checking for available code but also rating its ease of reproduction.\footnote{The project's database has since been extended to also include SIGGRAPH 2019--2021 as well as SIGGRAPH Asia 2015, at least partially.} While they found an overall increase of papers with code from $\approx$\,30\% in 2014 to $>$50\% in 2018, at the time of publication only 5 papers had a GRSI-certified reproducibility, yet 4 of these with a top reproducibility rating. For the visualization field, Haroz \cite{Haroz:2018:OPV} analyzed IEEE VIS 2017 papers\footnote{Later extended to IEEE VIS 2018--2019: \href{http://oavis.org/}{\texttt{oavis.org}}.} and, among other things, checked for the availability of source code that was used for data collection, the availability of the raw collected data, and the availability of materials or source code needed to reproduce the data analysis, with overall low results. Some notable exceptions with good reproducibility were highlighted by Kosara and Haroz \cite{Kosara:2018:SRC} in the same year. Looking at GRSI-certifications in 2020, Fekete and Freire \cite{Fekete:2020:ERV} only found few visualization papers with a GRS and call for more reproducibility and replicability in the field, in particular within the GRSI. Overall, however, few if any reliable numbers on how much visualization paper authors emphasize on reproducibility exist at this point---which is what I add with my analysis of GRSI-awarded papers in this work.

But even with the few numbers on the support of reproducibility in our field that have been published so far we already know that we need to do better. Several groups of authors have discussed ways to improve the situation for various visualization subfields \cite{Chopra:2023:PGG,Cushing:2018:SVR,Sukumar:2018:TDU,Valdez:2018:RRR} or how to include the subject into our taught curricula \cite{Syeda:2024:VRT}. More generally, Reina \cite{Reina:2023:CID} recommends and shows how to embed all information needed to reproduce a visual result (image) into the image itself. Another way that does not require source code is to provide executable demonstrators \cite{Bonneel:2020:CRC,Isenberg:2022:PEP}, which avoid, in particular, the compilation issues of code \cite{Bonneel:2020:CRC}. Other than the replicability of specific approaches, Garkov et al. \cite{Garkov:2022:RDC} argue that also the curation and long-term management of dedicated research datasets and artifacts can also be a means to facilitate research repeatability because, for instance, both ensure that new approaches can easily be compared to older results. Fekete and Freire \cite{Fekete:2020:ERV} discuss possibilities to ensure reproducibility specifically tailored to the different contribution types we have in the visualization field, while Franke et al. \cite{Franke:2023:TRV} offer a typology of reproducibility aspects that allows paper authors to decide on reproducibility and replicability aspects based on their specific situation. There also used to be the EuroRV\textsuperscript{3} workshop\footnote{\href{https://diglib7.eg.org/handle/10.2312/980}{\texttt{diglib7\discretionary{}{.}{.}eg\discretionary{}{.}{.}org\discretionary{/}{}{/}handle\discretionary{/}{}{/}10\discretionary{}{.}{.}2312\discretionary{/}{}{/}980}}} as dedicated venue to discuss the topic, yet its last incarnation happened as TrustVis five years ago in 2019 and not again since then.

I want to emphasize that there are many aspects of the larger open science movement (\eg, \cite{Jansen:2024:MWB}) that are important in this context. While work that focuses on empirical studies, for example, may require a certain type of shared material \cite{Kosara:2018:SRC} (\eg, pre-registrations or registered reports), in this paper I am looking generally at the issue of to what degree reproducibility of some sort of source code aspects (be it code to produce visual results or code to reproduce the statistical analysis of experimental data from a user experiment) is actually done in our field using the GRSI as the only certification authority available to our community at the moment.

\section{Data acquisition}
\label{sec:data}

To do so, I started by getting the data about reproducible papers from the listing on \href{https://www.replicabilitystamp.org/}{\texttt{replicabilitystamp\discretionary{}{.}{.}org}}, which---as of \GrsiDataCurrentAsOf---lists \GrsiTotalPapers{} papers in total. Using Python code with the BeautifulSoup library, I extracted each paper's title, its authors, the journal, the DOI, and the active and archived repository locations. I then sanitized the downloaded data by, in particular, correcting author names as much as possible (needed due to typos, missing special characters, first name last name order, name changes from marriages, and other errors), resulting in the identification of \GrsiTotalAuthors{} unique authors. For each paper I then queried the digital libraries (\ie, IEEE Xplore, Elsevier, ACM, or Crossref) to obtain publication data (publication year, volume, number, pages, etc.) and I manually added the authors' respective country of affiliation based on what authors reported on the published papers (often effective at the time of submission or publication; if multiple countries were named I added them all).

Next, I classified each paper as being a visualization paper or not. Because a manual or keyword-based classification is inherently biased, I attempted to use the paper authors' own classification as much as possible. I thus used data about whether a paper was \emph{presented} in a clearly visualization-focused venue, \ie, proper (journal-level) IEEE VIS\discretionary{/}{}{/}Euro\-Vis\discretionary{/}{}{/}Paci\-fic\-Vis papers,\footnote{The conference data comes from the VisPubData dataset \cite{Isenberg:2017:VMC} (for IEEE VIS), from the Eurographics digital library (for EuroVis; including both regular papers and STAR papers), or was extracted from the conference webpage (for PacificVis).} journal papers presented at IEEE VIS/EuroVis/PacificVis, etc.,\footnote{\label{foot:vis_presentations}I extracted this data largely from the respective conference webpages or programs. The data of regular papers and presentations at IEEE VIS is updated up to the conference's 2024 edition courtesy of Ross Maciejewski, Lane Harrison, and the VIS Papers Chairs, but the IEEE VIS 2024 data is preliminary at this point and there may be some changes as the final program for IEEE VIS 2024 is being finished (\ie, for the 2024 values in \autoref{fig:teaser}, in \autoref{fig:vis-grs-presentations-by-year}, and in \autoref{fig:vis-grs-presentations-by-year-linegraph}).} or if the papers were published in a journal special issue focused on visualization topics (\eg, special-issue versions of VCBM, EuroVA, MolVA, EnvirVis in Elsevier's C\&G).\footnote{\label{foot:c-and-g-special}I extracted this data from the special issue classifications in Elsevier's digital library and from special issue overview articles in C\&G. Please note that the 2023 and 2024 values of these types of publications are not yet final (\ie, the values for the bars in \autoref{fig:teaser}, in \autoref{fig:vis-grs-presentations-by-year}, and in \autoref{fig:vis-grs-presentations-by-year-linegraph} are too low or do not exist yet), as they are still being assembled.} If this failed, I also checked the titles of the remaining papers for a number of visualization-related keywords (\GrsiVisByKeywordPapersCount{}\texttimes; see the list of keywords in \autoref{app:keywords}). Finally, I also manually classified a small number of papers based on an inspection (\GrsiVisManuallyMarkedPapersCount{}\texttimes; see \autoref{app:manual-reasons} for the rationales). As noted, these last two ways of classification are arguably biased, but it affected a relatively low number of papers (\GrsiVisKeywordPlusManualPapersCount{} out of \GrsiTotalVisPapers{} papers classified as visualization papers in total, or \GrsiVisKeywordPlusManualPapersPercentage{}\%). This process resulted in \GrsiPercentageVisPapers{}\% of all GRSI-awarded papers being classified as discussing visualization topics (\GrsiTotalVisPapers{} out of \GrsiTotalPapers{}), with \GrsiTotalVisAuthors{} authors out of \GrsiTotalAuthors{} being classified as visualization authors (\ie, $\geq$50\% of their papers being classified as visualization papers).

\section{A visual GRSI data analysis}
\label{sec:visual_analysis}

\begin{figure}
	\centering
	\includegraphics[width=\linewidth]{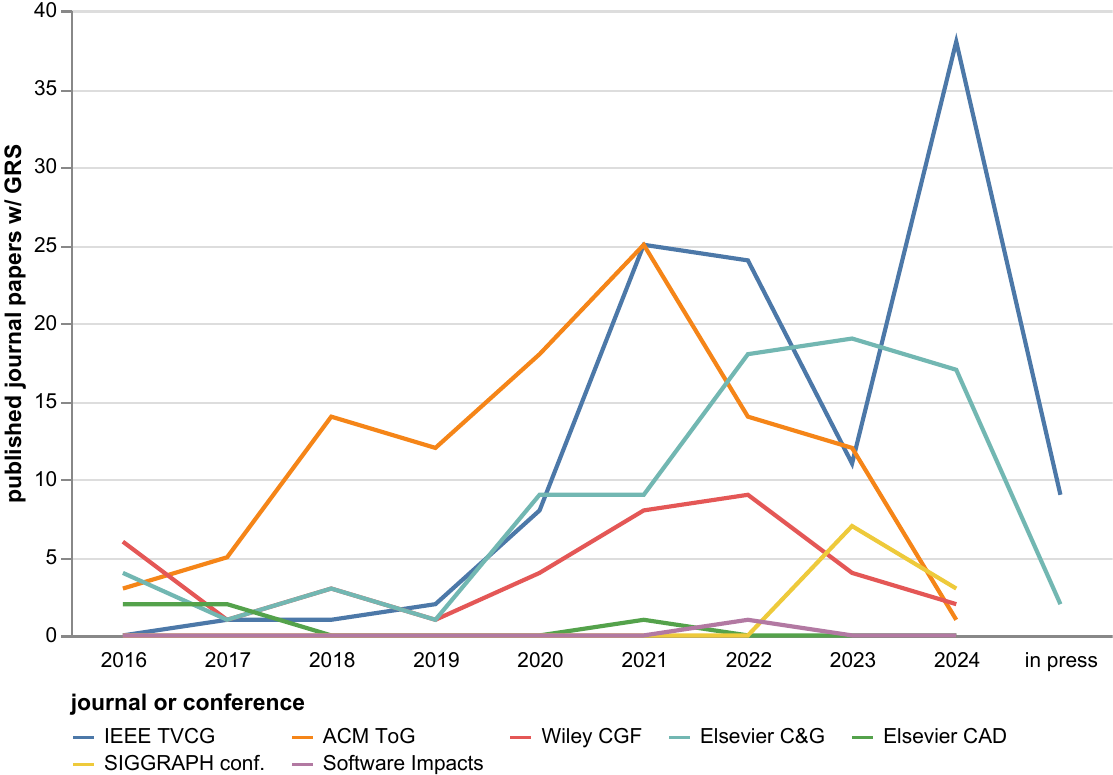}
	\caption{Overall development of papers with GRS, by publication venues (and their article \emph{publication} years). For a bar chart version see \autoref{fig:grs-overall-bars} in \autoref{app:plots}.}
	\label{fig:grs-overall}
\end{figure}

\begin{figure}
	\centering
	\includegraphics[width=\linewidth]{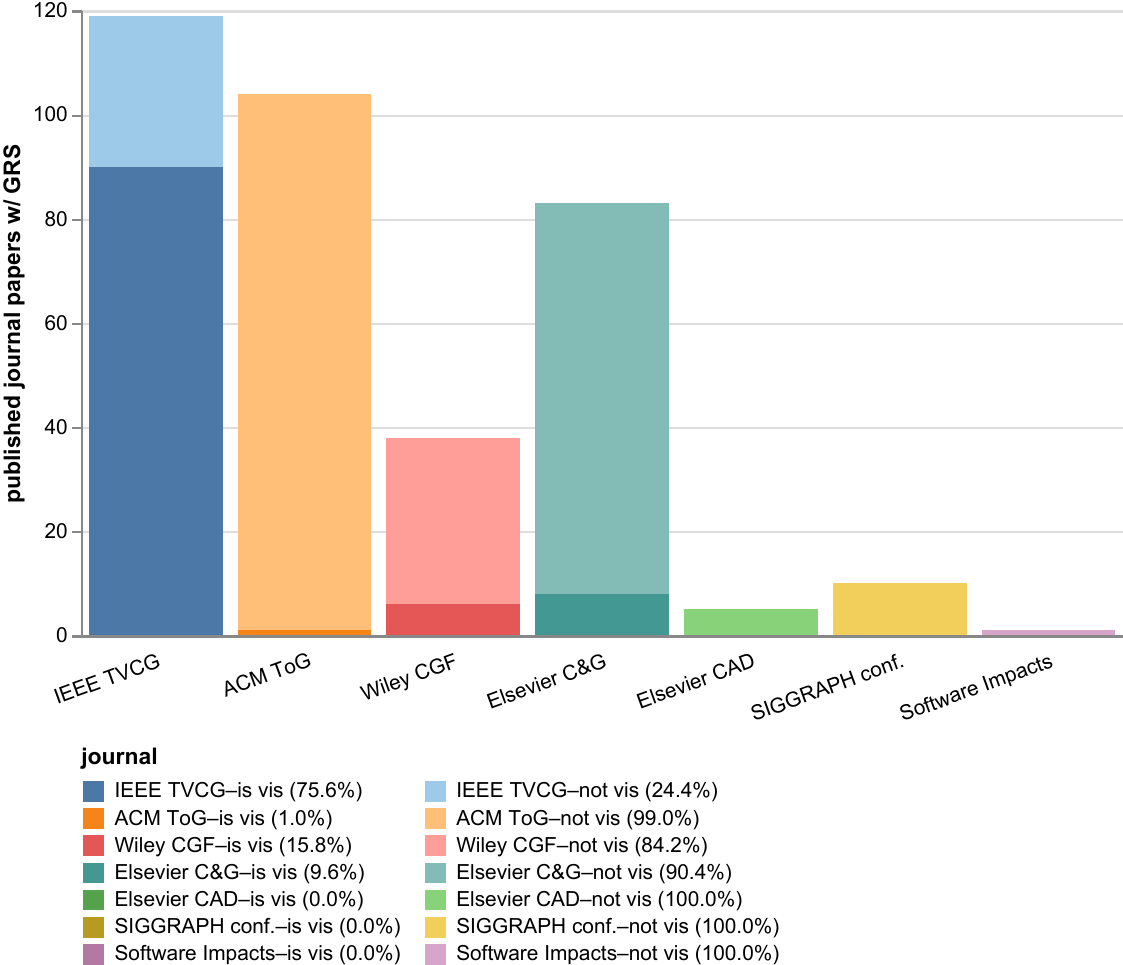}
	\caption{GRS papers by publication venue and their respective proportion classified as visualization papers. For a normalized version of the plot see \autoref{fig:grs-split-vis-novis-stackedbar-normalized} in \autoref{app:plots}.}
	\label{fig:grs-split-vis-novis-stackedbar}
\end{figure}

\begin{figure}
	\centering
	\includegraphics[width=\linewidth]{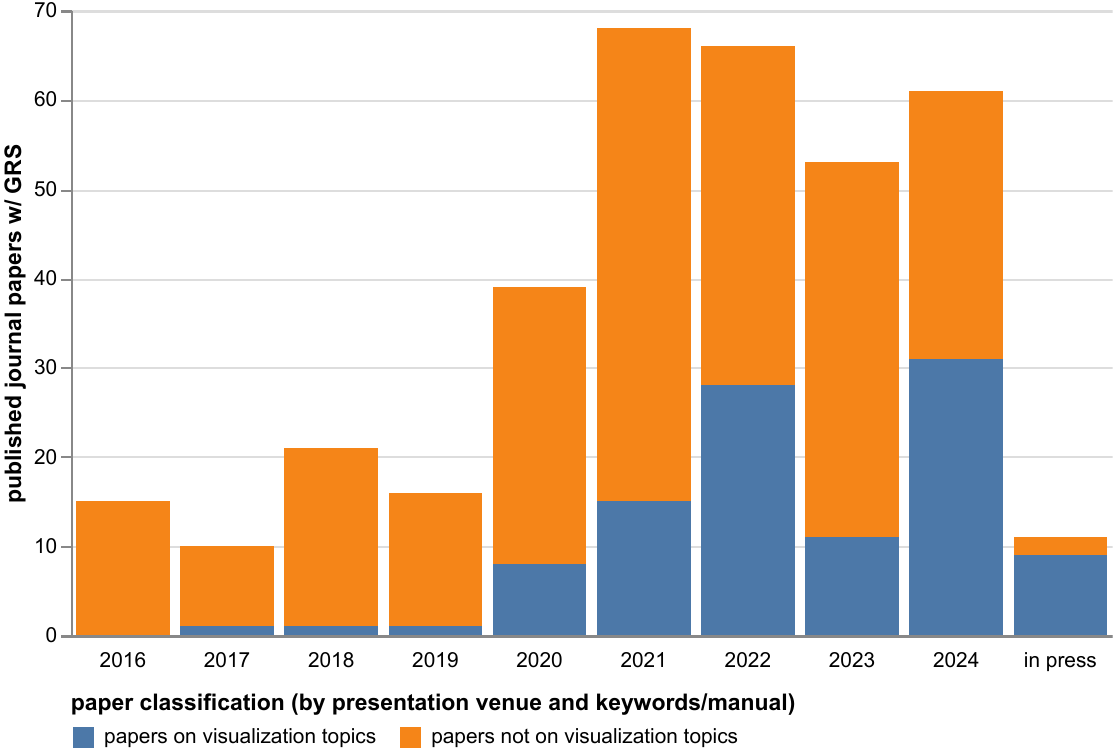}
	\caption{Overall proportion of GRS papers classified as visualization papers by article \emph{publication} years. For a normalized version of the plot see \autoref{fig:grs-split-vis-novis-normalized} in \autoref{app:plots}.}
	\label{fig:grs-split-vis-novis}
\end{figure}

I begin by looking at the temporal development of GRSI-awarded papers overall.\footnote{\label{foot:grsi-statistics}Unlike the statistics on the GRSI website (\href{https://www.replicabilitystamp.org/statistics.html}{\texttt{replicabilitystamp\discretionary{}{.}{.}org\discretionary{/}{}{/}statistics\discretionary{}{.}{.}html}}), I look at the temporal evolution of the \emph{publication} data, not when a paper was awarded the GRS.} \autoref{fig:grs-overall} shows that the first stamps were awarded for papers published in 2016, with the major contributors being the \emph{ACM Transactions on Graphics} (ToG) and the \emph{IEEE Transactions on Visualization and Computer Graphics} (TVCG), followed by Elsevier's \emph{Computers \& Graphics} (C\&G) and Wiley's \emph{Computer Graphics Forum} (CGF). \autoref{fig:grs-split-vis-novis-stackedbar} shows that TVCG and ToG are at about the same level, with C\&G and CGF having lower numbers, the remaining journals and the recently added SIGGRAPH conference papers do not (yet) play a big role. While it is normal that more recent years have lower numbers that those volumes a bit older (see also the total paper trend in \autoref{fig:grs-split-vis-novis}), what's interesting about \autoref{fig:grs-overall} is the distinct drop in numbers for TVCG in 2023 and the following rise in 2024. The likely reason is that in 2023, for some reason, the promise by TVCG\footnote{\label{foot:tvcg-fast-track}This benefit is promised by TVCG in their paper acceptance messages.} to assign a GRSI-awarded paper to the next available issue of the journal somehow did not get upheld, and only after a friendly reminder to TVCG did several ``in press'' articles get fully published in early 2024. As such the 2023 and 2024 numbers for TVCG are likely to be outliers---a view that is supported by the more or less constant or even continuously growing submission and acceptance numbers for TVCG articles as it is evident in the online stamp award time statistics of the GRSI.\textsuperscript{\ref{foot:grsi-statistics}}

\begin{figure}
	\centering
	\includegraphics[width=\linewidth]{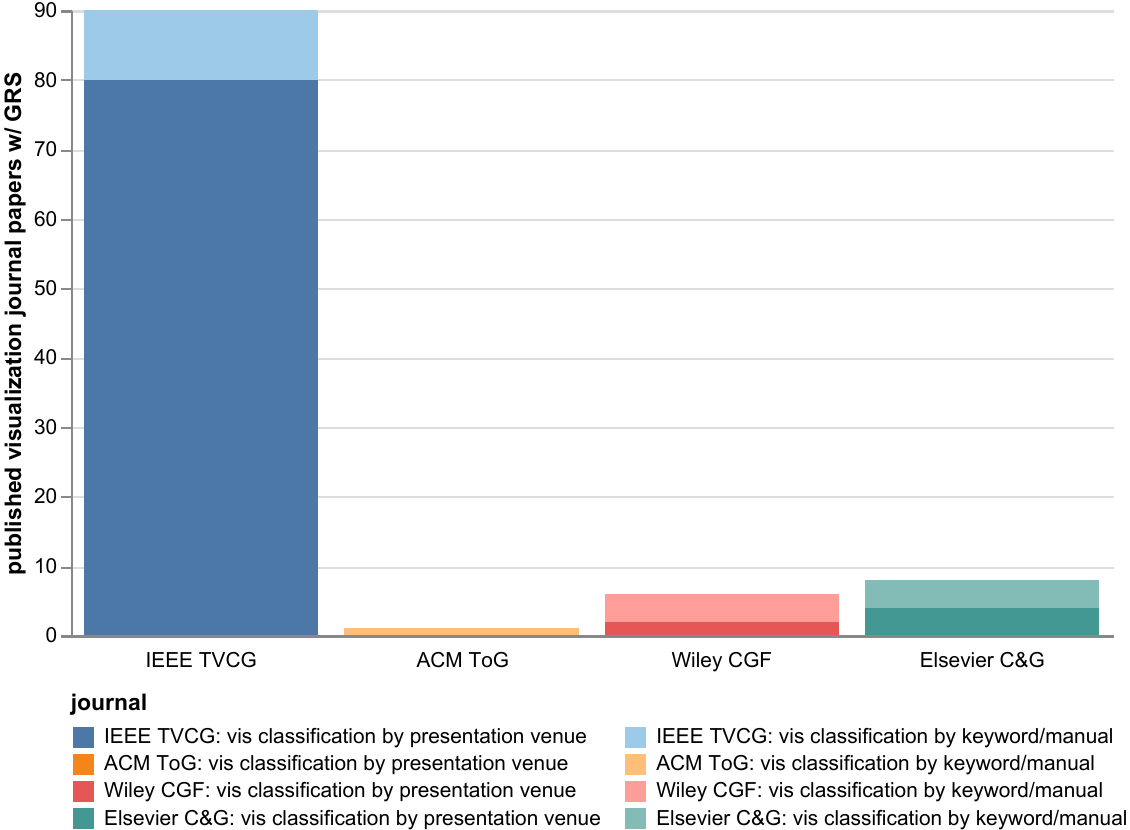}
	\caption{Sources of visualization papers overall, split between classification by visualization presentation venue and keyword/manual classification. A normalized version of this plot can be found in \autoref{fig:vis-grs-journals-aggregated-normalized} in \autoref{app:plots}.}
	\label{fig:vis-grs-journals-aggregated}
\end{figure}

\begin{figure}
	\centering
	\includegraphics[width=\linewidth]{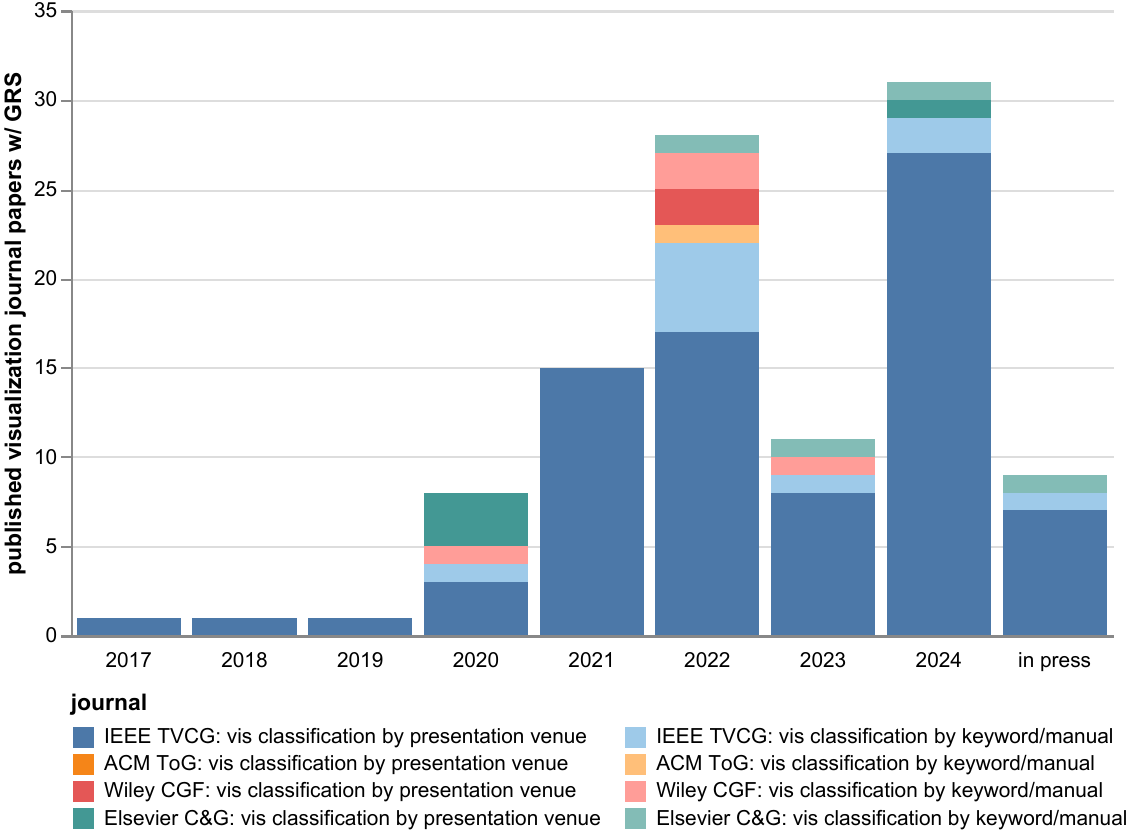}
	\caption{Same data as in \autoref{fig:vis-grs-journals-aggregated}, but by article \emph{publication} year. For a normalized version of the plot see \autoref{fig:vis-grs-journals-by-year-normalized} in \autoref{app:plots}.}
	\label{fig:vis-grs-journals-by-year}
\end{figure}

Looking next at the development of visualization contributions among all GRSI-awarded papers, in \autoref{fig:grs-split-vis-novis} we can see that in the first four years only few if any papers were classified as covering visualization topics---likely an effect from the GRSI originating from the computer graphics field. TVCG as the prime journal of the visualization field only joined the GRSI in 2018, and in \autoref{fig:vis-grs-journals-aggregated} we see that TVCG is by far the dominating source of GRSI-awarded visualization papers. This figure also shows the split between author-classified (\ie, classified by presentation venue) and keyword- or manually classified papers separately for each journal, with \GrsiVisPapersInIEEETVCGPercentagePresentation{}\% (\GrsiVisPapersInIEEETVCGPresentation{} out of \GrsiVisPapersInIEEETVCGTotal) of GRSI-awarded TVCG papers on visualization topics actually being presented by their authors at visualization conferences. The percentages at the other venues are much lower, with \GrsiVisPapersInWileyCGFPercentagePresentation{}\% (\GrsiVisPapersInWileyCGFPresentation{} out of \GrsiVisPapersInWileyCGFTotal) for CGF, \GrsiVisPapersInElsevierCaGPercentagePresentation{}\% (\GrsiVisPapersInElsevierCaGPresentation{} out of \GrsiVisPapersInElsevierCaGTotal) for C\&G, and \GrsiVisPapersInACMToGPercentagePresentation{}\% (\GrsiVisPapersInACMToGPresentation{} out of \GrsiVisPapersInACMToGTotal) for ToG (albeit at much lower total visualization papers in these three journals). For more detail, \autoref{fig:vis-grs-journals-by-year} shows the same data as \autoref{fig:vis-grs-journals-aggregated}, but with the development over the years. What is interesting to observe here is that almost all visualization TVCG papers are actually presented at conferences by their authors, except of those published in 2022---for which I really cannot envision a meaningful reason, other than it being a fluke due to the small-number statistics that this analysis admittedly is.

\begin{figure}
	\centering
	\includegraphics[width=\linewidth]{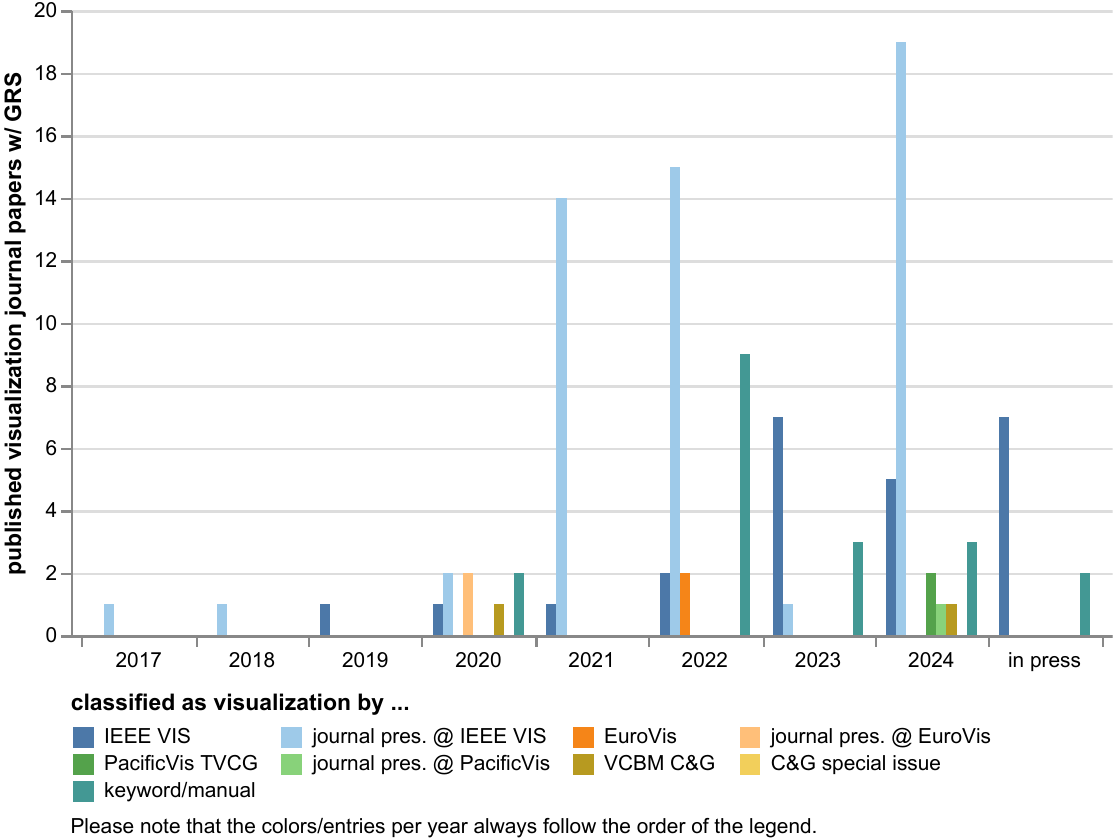}
	\caption{Different types of classification of papers as visualization work, by article \emph{publication} year (\ie, not by paper \emph{presentation} year). For a stacked bar chart version of the plot and its normalized version see \autoref{fig:grs-vis-classification-barchart} and \autoref{fig:grs-vis-classification-barchart-normalized} in \autoref{app:plots}, respectively.}
	\label{fig:grs-vis-classification}
\end{figure}

\begin{figure*}
	\centering
	\includegraphics[width=\linewidth]{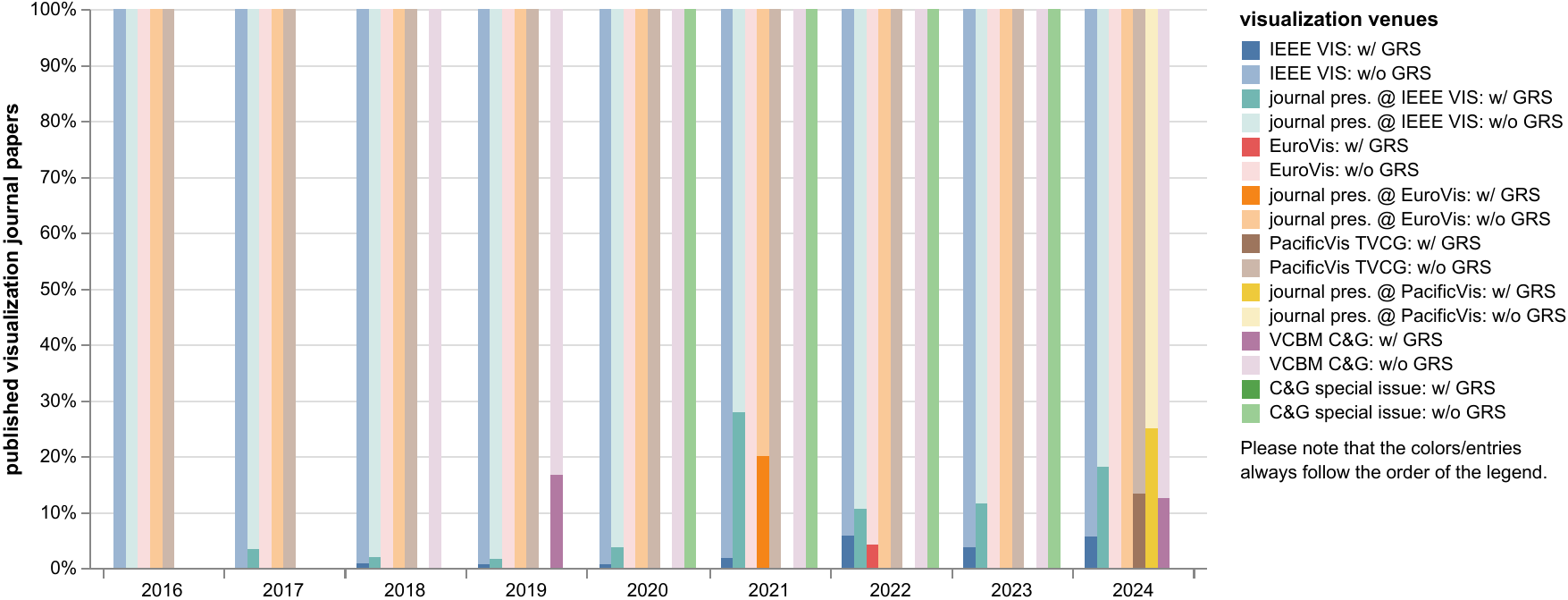}
	\caption{Papers that can clearly be classified as visualization work (based on being accepted to visualization conferences, being journal presentations at visualization conferences,\textsuperscript{\ref{foot:vis_presentations},\ref{foot:c-and-g-special}} or having appeared in visualization-themed special issues in journals) by \emph{presentation} year (\ie, based on the \emph{conference years}, in which the papers were presented; same data as in \autoref{fig:teaser}), with their GRS proportions, normalized.}
	\label{fig:vis-grs-presentations-by-year}
\end{figure*}

This small-number character is even more apparent if we drill deeper into the different venues where each of the visualization papers was presented. \autoref{fig:grs-vis-classification} shows the evolution of GRSI-awarded visualization papers by their \emph{publication} years, split up into the different conferences at which they were accepted (\GrsiIeeeVisPapersCount{} at IEEE VIS, \GrsiEuroVisPapersCount{} at EuroVis, and \GrsiPacificVisTvcgPapersCount{} at PacificVis), the conferences where the papers were presented as journal papers (\GrsiIeeeVisJournalPresentationsCount{} at IEEE VIS, \GrsiEuroVisJournalPresentationsCount{} at EuroVis, and \GrsiPacificVisJournalPresentationsCount{} at PacificVis), special-issue journal papers (\GrsiVcbmCagPapersCount{} in C\&G special issue on VCBM, \GrsiCagSpecialIssuesPapersCount{} in a C\&G special issue on EuroVA, EnvirVis, or MolVA), or the \GrsiVisKeywordPlusManualPapersCount{} keyword- or manually classified papers. \autoref{fig:teaser} shows the same data, but focuses only on the conferences and the papers presented in their programs (regular and journal presentations) and also shows the papers by the year in which they were \emph{presented}. Please note that in this figure (and in the normalized version in \autoref{fig:vis-grs-presentations-by-year}) the IEEE VIS journal presentation counts include the IEEE Computer Graphics and Applications (CG\&A) papers presented at the conference. This magazine, however, is not (yet?) included within the scope of venues for which a GRS is awarded, so that currently even in an ideal case the IEEE VIS journal papers could not achieve a 100\% GRS rate (and also the presented rate is thus lower than it actually is for only presented TVCG papers).

Nonetheless, while we see in \hyperref[fig:teaser]{Figures \ref{fig:teaser}} and~\ref{fig:vis-grs-presentations-by-year} that the overall number of GRSI-awarded papers within visualization venues is still low, it is increasing. In particular for journal papers presented at IEEE VIS we see a substantial number of papers with a GRS---\GrsiIeeeVisJournalPresentationsCount{} papers in total so far---, leading to percentages of $\approx$\,10\% to up to $\approx$\,20\%--30\% in recent years.
For instance, among the TVCG papers to be presented at IEEE VIS 2024, \percentageRounded{\GrsiIeeeVisTVCGJournalPapersInMMXXIV}{\TotalIeeeVisTVCGJournalPapersInMMXXIV}\% are already certified at this point, \percentageRounded{\GrsiIeeeVisTVCGJournalPapersInMMXXIII}{\TotalIeeeVisTVCGJournalPapersInMMXXIII}\% for the TVCG papers that were presented at VIS in 2023, \percentageRounded{\GrsiIeeeVisTVCGJournalPapersInMMXXII}{\TotalIeeeVisTVCGJournalPapersInMMXXII}\% for 2022, and \percentageRounded{\GrsiIeeeVisTVCGJournalPapersInMMXXI}{\TotalIeeeVisTVCGJournalPapersInMMXXI}\% for 2021. 
This is great news, and the fact that the percentage of GRSI-awarded papers among the TVCG journal papers is much higher than the percentage among pure IEEE VIS papers (\percentageRounded{(\GrsiIeeeVisPapersInMMXXIV+\GrsiIeeeVisPapersInMMXXIII+\GrsiIeeeVisPapersInMMXXII+\GrsiIeeeVisPapersInMMXXI)}{(\TotalIeeeVisPapersInMMXXIV+\TotalIeeeVisPapersInMMXXIII+\TotalIeeeVisPapersInMMXXII+\TotalIeeeVisPapersInMMXXI)}\% on average in the period 2021--2024) is also not surprising: pure IEEE VIS papers are guaranteed publication in (typically) the first TVCG issue of the year following the conference (\ie, $\approx$\,6 months after paper acceptance and $\approx$\,3 months after presentation), while non-con\-fe\-rence TVCG papers normally have to wait 1--2 years for full publication (\ie, being assigned a volume, an issue, and page numbers) if it were not for TVCG's offer to fast-track GRSI-awarded papers to full publication.\textsuperscript{\ref{foot:tvcg-fast-track}}

\begin{figure}
	\centering
	\includegraphics[height=.36\linewidth]{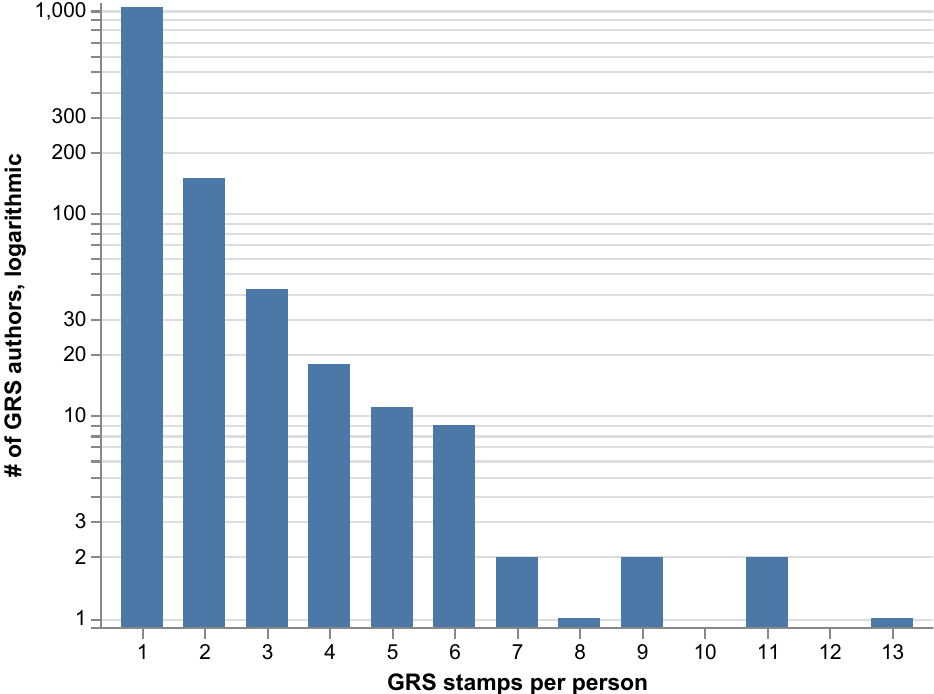}\hfill%
	\includegraphics[height=.36\linewidth]{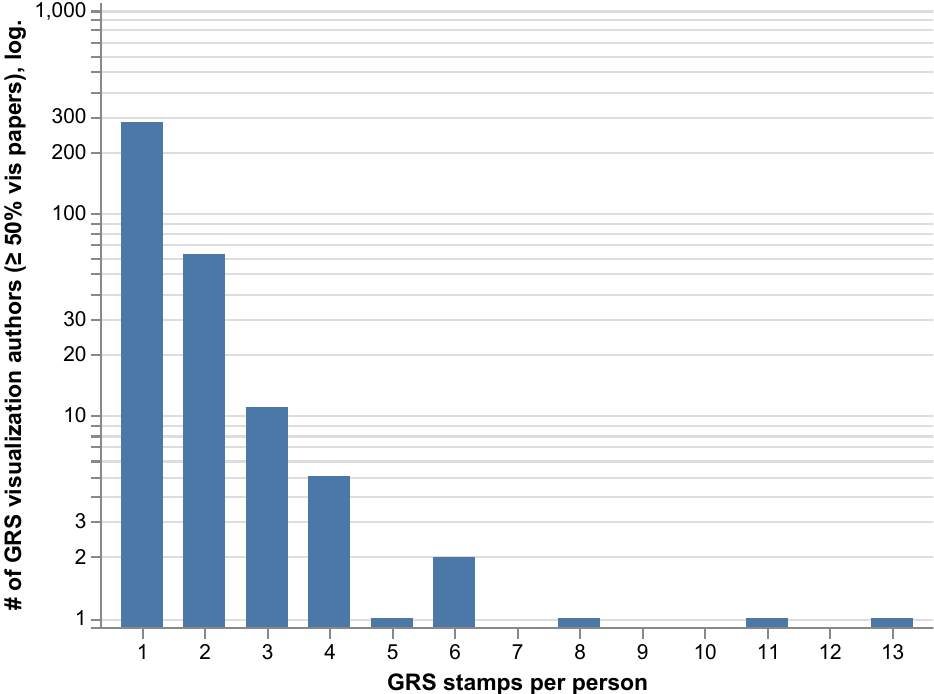}%
	\caption{Number of authors with a given number of GRSI-awarded papers. Left: for all authors; right: only visualization authors.}
	\label{fig:vis-grs-per-author}
\end{figure}

\begin{figure}
	\centering
	\includegraphics[height=.39\linewidth]{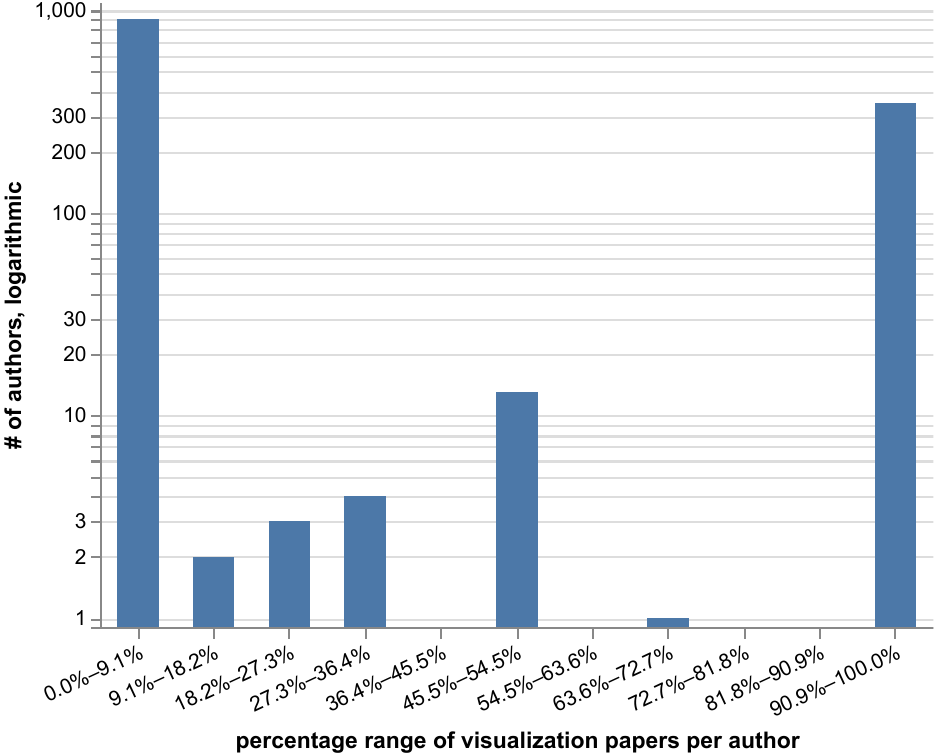}\hfill%
	\includegraphics[height=.39\linewidth]{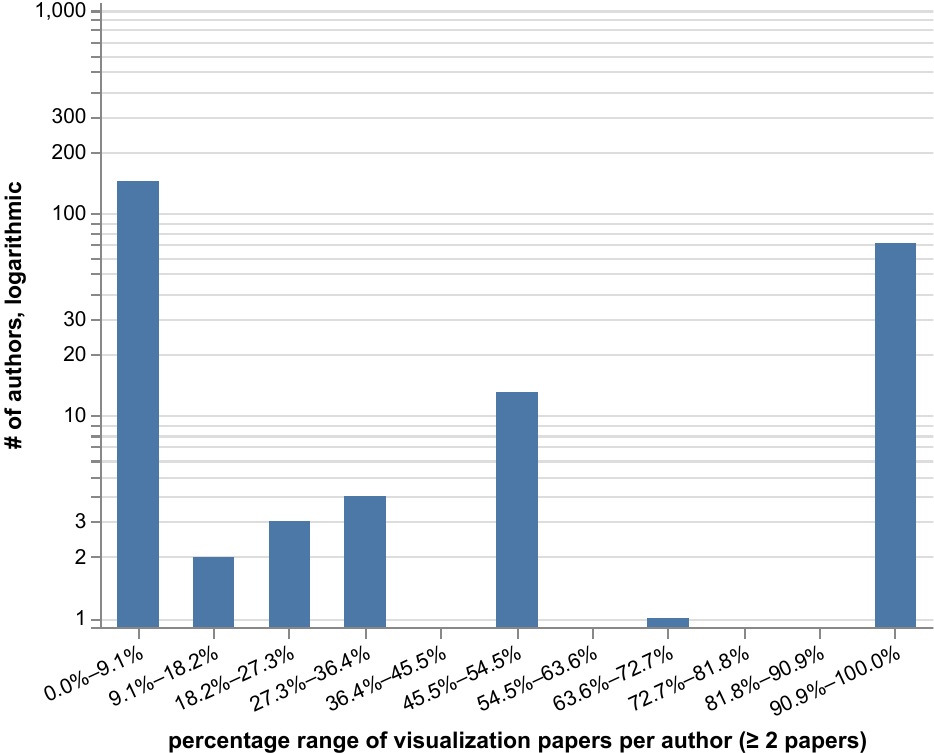}%
	\caption{Histogram of the percentage of visualization papers per author (logarithmically plotted; left: all authors, right: only authors with $\geq$ 2 papers): a fairly bimodal distribution.}
	\label{fig:histogram-author-vis-percentage}
\end{figure}

The data cleaning and visualization classification now also allows us to look at the contributions to the GRSI by author. \autoref{fig:vis-grs-per-author} shows the overall number of authors for a given number of GRSI awards---on the left for all authors who received a GRS, on the right only for people I classify as visualization authors by the fact that they, in my analysis, published $\geq$50\% visualization papers. We see the clear signs of a power-law distribution, which is to be expected for datasets like this one. With our data, however, we can also ask to what degree the different communities between visualization and computer graphics intermix (I assume that papers not about visualization topics can be classified as being about computer graphics and related topics). \autoref{fig:histogram-author-vis-percentage} thus shows a histogram of number of authors with a given percentage of visualization papers (based purely on the GRSI data). The left histogram in \autoref{fig:histogram-author-vis-percentage} shows this analysis for all authors, but naturally this plot is biased by the many authors with a single paper only in the data (as we saw in \autoref{fig:vis-grs-per-author})---these can by definition only contribute to either group exclusively. On the right of \autoref{fig:histogram-author-vis-percentage} I show, therefore, only those authors with $\geq$2 papers in the dataset. Still, we can see that we have a rather bimodal distribution between visualization authors and computer graphics authors, with only few authors publishing (GRSI-awarded) papers in both fields.

\begin{figure}
	\centering
	\includegraphics[height=.45\linewidth]{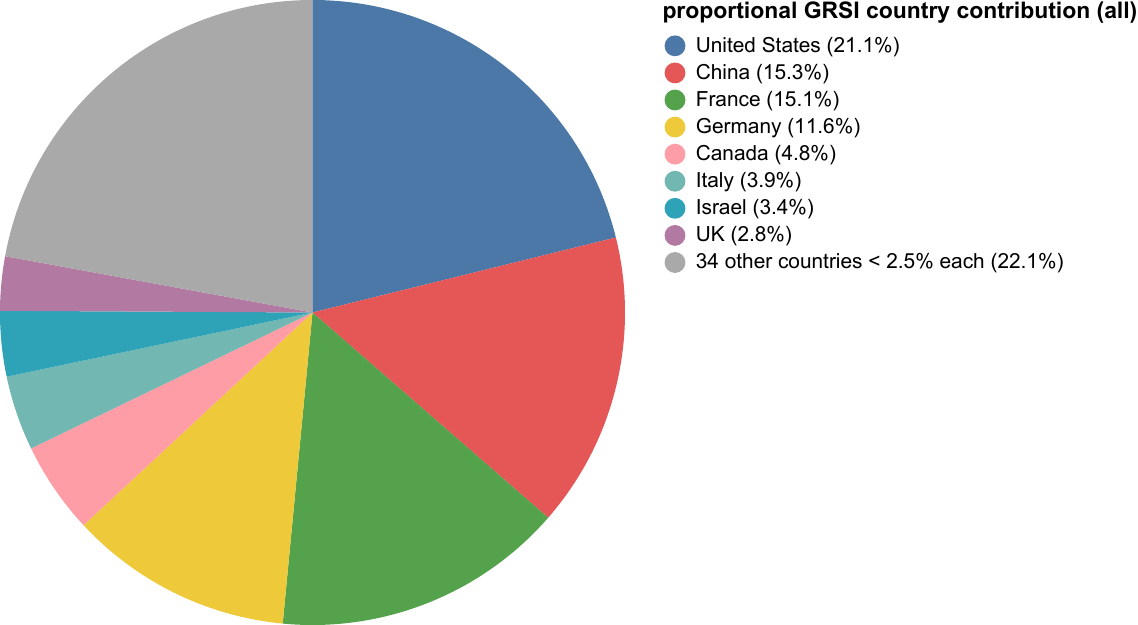}\\[1ex]%
	\includegraphics[height=.45\linewidth]{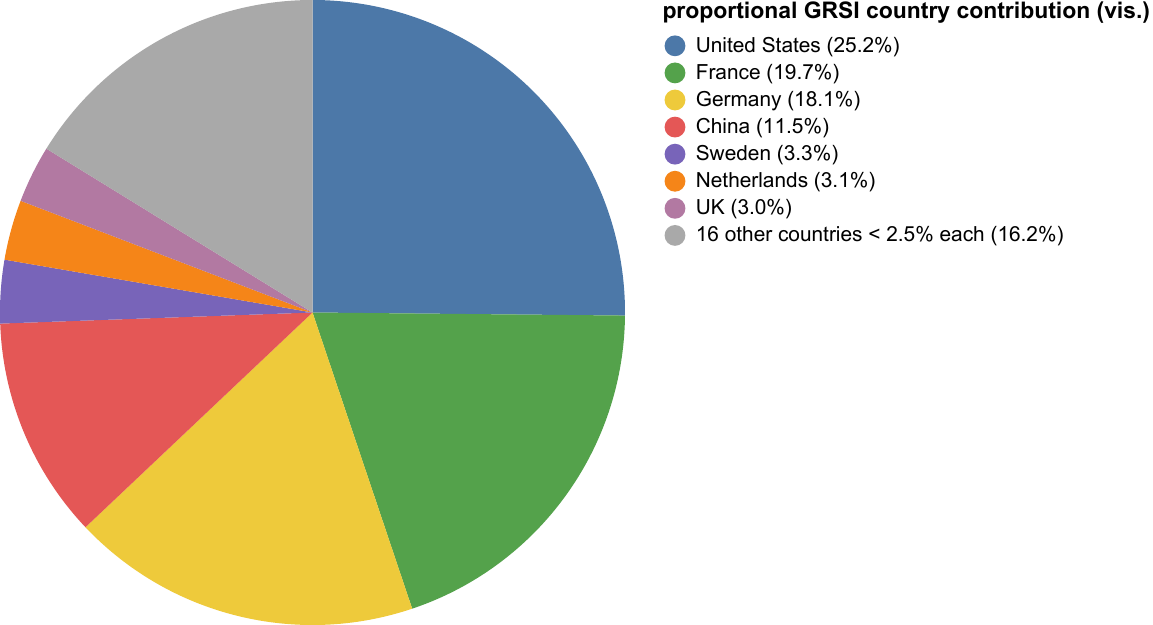}%
	\caption{Proportional country contribution to GRS awards (counting all paper authors). Top: overall; bottom: visualization papers. These plots are thresholded to \GrsiCountryPieChartThreshold\% per-country contribution (for clarity), for full versions see \autoref{fig:grs_all-per-country-comparison-proportional-and-absolute} and \autoref{fig:grs_visualization-per-country-comparison-proportional-and-absolute} in \autoref{app:plots}, respectively.}
	\label{fig:vis-grs-per-country-proportional}
\end{figure}

\begin{figure}
	\centering
	\includegraphics[height=.45\linewidth]{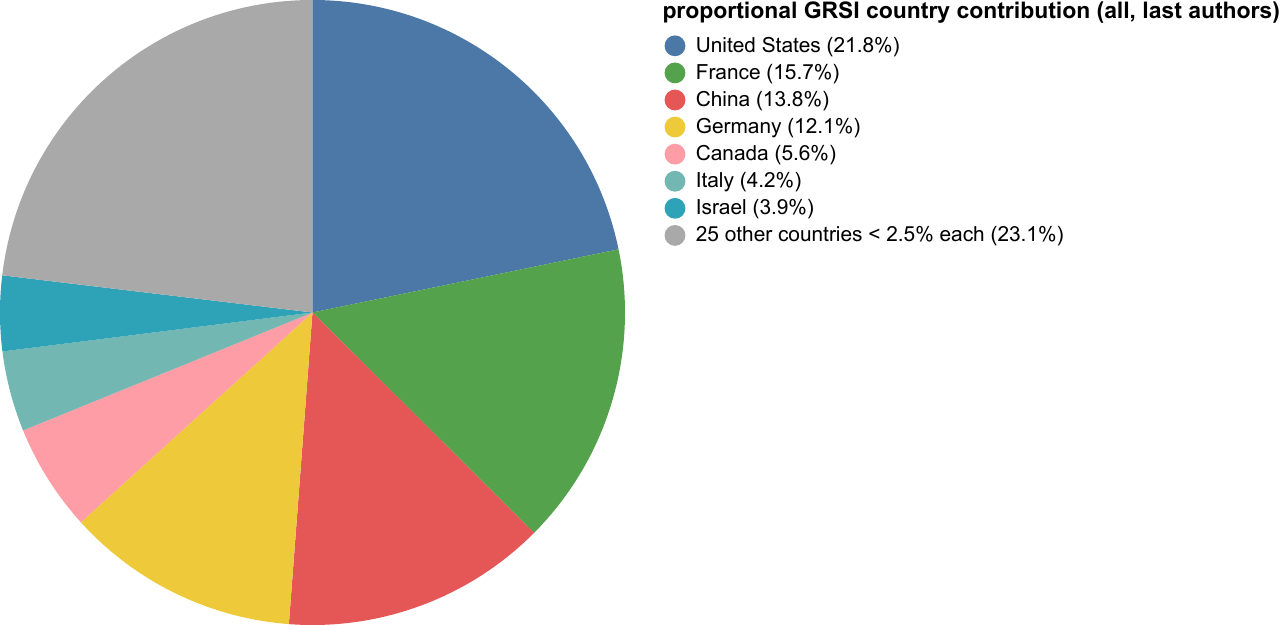}\\[1ex]%
	\includegraphics[height=.45\linewidth]{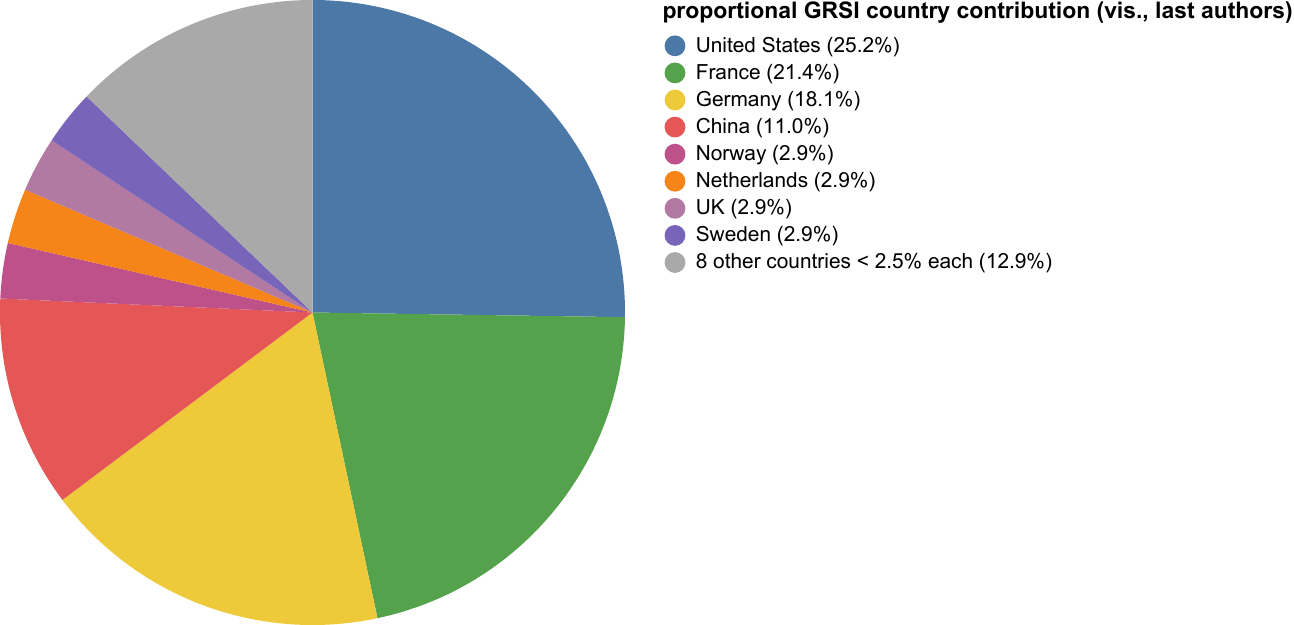}%
	\caption{Country contribution of paper last authors to GRS awards. Top: overall; bottom: visualization papers. These plots are thresholded to \GrsiCountryPieChartThreshold\% per-country contribution (for clarity), for full versions see \autoref{fig:grs_all_senior-per-country-comparison-proportional-and-absolute} and \autoref{fig:grs_visualization_senior-per-country-comparison-proportional-and-absolute} in \autoref{app:plots}, respectively.}
	\label{fig:vis-grs-per-country-proportional-last-authors}
\end{figure}

Finally, we can investigate the countries in which the GRSI-awarded authors are active. I base this analysis on the author affiliation country or countries at publication time (as they reported it on the paper), as I described in \autoref{sec:data},\ifthenelse{0<\GrsiDifferenceInPaperDatabases}{\footnote{Due to the way the data analysis is implemented, the dataset of countries of affiliations currently has \GrsiDifferenceInPaperDatabases{} \ifthenelse{1<\GrsiDifferenceInPaperDatabases}{entries}{entry} fewer than that of the overall papers. \ifthenelse{1<\GrsiDifferenceInPaperDatabases}{These papers currently do}{This one paper currently does} not have a DOI assigned as or right now, so that I cannot (yet) get the official data from the corresponding digital library.}}{} and for the purpose of the analysis I assign each paper the same weight of 1. I then distribute this weight evenly to each author of the paper, so an author of a paper with $N$ co-authors gets weight \nicefrac{1}{$N$}, which is then assigned to the respective country of affiliation of the author. If an author has $M$ affiliations in different countries, then I only award each of these countries a weight of \nicefrac{1}{$N \cdot M$} from the author. This way each paper counts equally, regardless of number of authors and their affiliations, and we get the proportional representation I show in \autoref{fig:vis-grs-per-country-proportional}. The top pie chart in the figure shows the countries for all GRSI-awarded authors, the bottom one shows the version for only visualization-themed papers. We can see that four countries play the major role in both cases. For all papers with GRS, \GrsiCountryPieChartOverallNoOneName{}{} has the largest proportion with \GrsiCountryPieChartOverallNoOnePercentage{}\%, followed by \GrsiCountryPieChartOverallNoTwoName{} with \GrsiCountryPieChartOverallNoTwoPercentage{}\%, \GrsiCountryPieChartOverallNoThreeName{} with \GrsiCountryPieChartOverallNoThreePercentage{}\%, and \GrsiCountryPieChartOverallNoFourName{} with \GrsiCountryPieChartOverallNoFourPercentage{}\%. For visualization-themed papers with GRS, \GrsiCountryPieChartVisNoOneName{} also has the largest proportion with \GrsiCountryPieChartVisNoOnePercentage{}\%, followed by \GrsiCountryPieChartVisNoTwoName{} with \GrsiCountryPieChartVisNoTwoPercentage{}\%, \GrsiCountryPieChartVisNoThreeName{} with \GrsiCountryPieChartVisNoThreePercentage{}\%, and \GrsiCountryPieChartVisNoFourName{} with \GrsiCountryPieChartVisNoFourPercentage{}\%. I also asked myself if this distribution would change drastically if we would only consider the last and thus likely the senior author of each paper, for which I show the analysis in \autoref{fig:vis-grs-per-country-proportional-last-authors}.\footnote{Behind this analysis is the assumption that the last\discretionary{/}{}{/}senior author is more likely to be the instigator of a GRSI application.} As we can see, the situation is similar to what we saw before. For all papers with GRS, \GrsiCountryPieChartOverallSeniorNoOneName{}{} again has the largest proportion with \GrsiCountryPieChartOverallSeniorNoOnePercentage{}\%, followed by \GrsiCountryPieChartOverallSeniorNoTwoName{} with \GrsiCountryPieChartOverallSeniorNoTwoPercentage{}\%, \GrsiCountryPieChartOverallSeniorNoThreeName{} with \GrsiCountryPieChartOverallSeniorNoThreePercentage{}\%, and \GrsiCountryPieChartOverallSeniorNoFourName{} with \GrsiCountryPieChartOverallSeniorNoFourPercentage{}\%. For visualization-themed papers with GRS, \GrsiCountryPieChartVisSeniorNoOneName{} also still has the largest proportion with \GrsiCountryPieChartVisSeniorNoOnePercentage{}\%, followed by \GrsiCountryPieChartVisSeniorNoTwoName{} with \GrsiCountryPieChartVisSeniorNoTwoPercentage{}\%, \GrsiCountryPieChartVisSeniorNoThreeName{} with \GrsiCountryPieChartVisSeniorNoThreePercentage{}\%, and \GrsiCountryPieChartVisSeniorNoFourName{} with \GrsiCountryPieChartVisSeniorNoFourPercentage{}\%.

In visualization we thus see that the mentioned four countries are responsible for $\approx$ \nicefrac{3}{4} of all the reproducible research. Of course, we would need to compare this number to the overall contributions of the different countries to the field, such as looking at all papers presented at IEEE VIS as the primary publication venue, to know if these numbers differ from the general situation. Unfortunately I do not have the data to check it, and augmenting the VisPubData dataset \cite{Isenberg:2017:VMC} with countries of affiliations would be a substantial amount of work---be\-yond the scope of this paper. My expectation, however, is that these numbers would differ. For example, the top three contributors to the GRSI from the visualization community (ranked 1\textsuperscript{st}, 2\textsuperscript{nd}, and 7\textsuperscript{th} overall), are from France at the moment, at least partially explaining the high ranking of the country in the analysis above. So at the moment and due to the low number of reproducible papers overall (when compared to the overall number of scientific publications in both visualization specifically and the larger computer graphics field in general) some countries may be over- or underrepresented, simply due to the initiative or lack of initiative of specific paper authors in the respective fields.

\section{Discussion}
\label{sec:discussion}

So how are we doing, then? Visualization papers currently only represent \GrsiPercentageVisPapers{}\% of all papers with a GRS, somewhere between a quarter and a third. And out of the \pgfmathparse{\TotalIeeeVisPapersInMMXXIV+\TotalIeeeVisPapersInMMXXIII+\TotalIeeeVisPapersInMMXXII+\TotalIeeeVisPapersInMMXXI}\pgfmathprintnumber[fixed, precision=0]{\pgfmathresult} full papers at IEEE VIS conferences 2021--2024, only \pgfmathparse{\GrsiIeeeVisPapersInMMXXIV+\GrsiIeeeVisPapersInMMXXIII+\GrsiIeeeVisPapersInMMXXII+\GrsiIeeeVisPapersInMMXXI}\pgfmathprintnumber[fixed, precision=0]{\pgfmathresult} (or \percentageRounded{(\GrsiIeeeVisPapersInMMXXIV+\GrsiIeeeVisPapersInMMXXIII+\GrsiIeeeVisPapersInMMXXII+\GrsiIeeeVisPapersInMMXXI)}{(\TotalIeeeVisPapersInMMXXIV+\TotalIeeeVisPapersInMMXXIII+\TotalIeeeVisPapersInMMXXII+\TotalIeeeVisPapersInMMXXI)}\%) have received a stamp as of today,\footnote{In total, \GrsiIeeeVisPapersCount{} IEEE VIS full papers have received a GRS to date.} and \pgfmathparse{\GrsiIeeeVisTVCGJournalPapersInMMXXIV+\GrsiIeeeVisTVCGJournalPapersInMMXXIII+\GrsiIeeeVisTVCGJournalPapersInMMXXII+\GrsiIeeeVisTVCGJournalPapersInMMXXI}\pgfmathprintnumber[fixed, precision=0]{\pgfmathresult} out of the \pgfmathparse{\TotalIeeeVisTVCGJournalPapersInMMXXIV+\TotalIeeeVisTVCGJournalPapersInMMXXIII+\TotalIeeeVisTVCGJournalPapersInMMXXII+\TotalIeeeVisTVCGJournalPapersInMMXXI}\pgfmathprintnumber[fixed, precision=0]{\pgfmathresult} TVCG journal presentations at IEEE VIS (or \percentageRounded{(\GrsiIeeeVisTVCGJournalPapersInMMXXIV+\GrsiIeeeVisTVCGJournalPapersInMMXXIII+\GrsiIeeeVisTVCGJournalPapersInMMXXII+\GrsiIeeeVisTVCGJournalPapersInMMXXI)}{(\TotalIeeeVisTVCGJournalPapersInMMXXIV+\TotalIeeeVisTVCGJournalPapersInMMXXIII+\TotalIeeeVisTVCGJournalPapersInMMXXII+\TotalIeeeVisTVCGJournalPapersInMMXXI)}\%) in the same time frame.\footnote{In total, \GrsiIeeeVisTvcgJournalPresentationsCount{} TVCG papers presented at VIS have received a GRS.} After all, there is a difference between, on the one hand, some code being available as open source for a paper (which we do see often in today's papers) that, with little or (usually much) more effort \cite{Bonneel:2020:CRC} can be made to run and, on the other hand, the actual certification of a sufficiently documented and runnable code---the latter being ensured by a GRSI certification. Research in psychology \cite{Crumwell:2023:WBC}, for instance, found that ``just'' awarding badges to papers for the fact that data and/or code is available is not sufficient to guarantee an exact reproducibility of the reported results---which, in contrast, the certification by initiatives such as the GRSI does guarantee (at least for those visuals of the paper that are being checked). Also within visualization we thus need more certification of reproducibility because only this certification forces authors to check that there shared code actually runs out of the box and has a sufficient level of documentation. The \GrsiPercentageVisPapers{}\%, the \percentageRounded{(\GrsiIeeeVisTVCGJournalPapersInMMXXIV+\GrsiIeeeVisTVCGJournalPapersInMMXXIII+\GrsiIeeeVisTVCGJournalPapersInMMXXII+\GrsiIeeeVisTVCGJournalPapersInMMXXI)}{(\TotalIeeeVisTVCGJournalPapersInMMXXIV+\TotalIeeeVisTVCGJournalPapersInMMXXIII+\TotalIeeeVisTVCGJournalPapersInMMXXII+\TotalIeeeVisTVCGJournalPapersInMMXXI)}\%, and especially the \percentageRounded{(\GrsiIeeeVisPapersInMMXXIV+\GrsiIeeeVisPapersInMMXXIII+\GrsiIeeeVisPapersInMMXXII+\GrsiIeeeVisPapersInMMXXI)}{(124+133+119+109)}\% are too low---we can and we should do better than that. Several calls for more reproducibility exist in our field as I have reviewed in \autoref{sec:rw} (\eg, \cite{Fekete:2020:ERV,Haroz:2018:OPV,Kosara:2016:EBS})---yet why are the numbers of papers at our main venues that have received a GRS still that low then? Speculating about possible reasons for this situation, a few reasons could explain the low numbers.

First, there may be an \textbf{issue with (too) few people knowing about the GRSI}. In this case we would need more publicity for this service. Yes, the EICs of the affected journals repeatedly mention the initiative in their messages to the reader (\eg, \cite{DeFloriani:2018:MEC,Jorge:2018:EN}), but maybe these messages are not as well read by the community? Also, while the GRSI is mentioned in the acceptance letter at least of some journals (such as TVCG) and conferences (such as IEEE VIS), maybe we could more clearly announce this initiative and its benefits for our community in the opening sessions of our conferences, so that more people hear about and are aware of it? On the bright side, however, as of today, \GrsiIeeeVisPapersInMMXXIV{} of the \TotalIeeeVisPapersInMMXXIV{} regular papers (\percentageRounded{\GrsiIeeeVisPapersInMMXXIV}{\TotalIeeeVisPapersInMMXXIV}\%) that are accepted to the (at the time of writing this paper still upcoming) 2024 VIS conference are already GRSI-certified---already more than the \GrsiIeeeVisPapersInMMXXIII{} papers (\percentageRounded{\GrsiIeeeVisPapersInMMXXIII}{\TotalIeeeVisPapersInMMXXIII}\%) of the 2023 conference 
listed to date. So at least some authors in our community are well aware of the GRSI and are eager to get their work certified, even before the conference.

Second, and despite this very positive observation just now, there may still be an \textbf{issue with few people willing to encourage the students to go the extra mile or the lack of tangible incentives}---after all, documentation of code still can require quite a bit of additional work, even if in many cases the GRSI application is rather lightweight. Yet being awarded a GRS currently often still boils down to ``eternal fame and glory'' \cite{Isenberg:2022:PEP}. One could argue (and some have \cite{Haroz:2018:OPV}) that a way to success is to make reproducibility a requirement for publication, but I am not sure how realistic such a move would be (at least in the short run), and we would need to deal with cases where code cannot be shared or where other reasons arguably prevent us from checking reproducibility. For instance, exceptions are possible even in models that require shared materials \cite{Haroz:2018:OPV} or the checking could be assigned to a trusted third party \cite{Besancon:2021:OSS}. Yet a better approach may be to begin with more strongly incentivizing reproducibility such as through reproducibility awards at our conferences and in our journals---beyond a GRSI certification. Computers \& Graphics already has a corresponding award, the ``Computers and Graphics Best (Replicable) Paper Award'' \cite{Jorge:2020:NEC} (\eg, \cite{Glencross:2021:GBP,Glencross:2022:GBP})---perhaps TVCG and CGF could introduce something similar as well? Maybe also other incentives could work, such as incentives at the level of the authors' local institutions that should award GRS-like achievements or provide other forms of credit for such efforts.

Third, there may be an \textbf{issue with visualization being underrepresented in the possible venues}. As we saw in \autoref{fig:grs-split-vis-novis-stackedbar}, the vast majority of GRSI-awarded visualization papers are published in TVCG. The other journals and conferences that are eligible for a GRS have little to no visualization-themed content. One way to address this issue would be to encourage the authors of visualization papers in those other venues to apply for the stamp, another is to add more core visualization venues to the list of venues to the GRSI. For instance, journals like Sage's \emph{Information Visualization},\footnote{\href{https://journals.sagepub.com/home/ivi}{\texttt{journals.sagepub.com/home/ivi}}} Springer's \emph{Journal of Visualization},\footnote{\href{https://link.springer.com/journal/12650}{\texttt{link.springer.com/journal/12650}}} IEEE's \emph{Computer Graphics and Applications},\footnote{\href{https://www.computer.org/csdl/magazine/cg}{\texttt{computer.org/csdl/magazine/cg}}} Elsevier's \emph{Visual Informatics},\footnote{\href{https://www.sciencedirect.com/journal/visual-informatics}{\texttt{sciencedirect.com/journal/visual-informatics}}} and \emph{The Journal of Visualization and Interaction}\footnote{\href{https://www.journalovi.org/}{\texttt{journalovi.org}}} could be added to the list. Also, similar to the GRSI recently having embraced SIGGRAPH's and SIGGRAPH Asia's conference-only papers, it could add at least some of the many significant smaller conferences and workshops of our field such as PacificVis (non-TVCG papers), ChinaVis, LDAV, VizSec, TopoInVis, VCBM (non-C\&G papers), IVAPP, plus many others, and last but not least also BELIV. If such an expansion of topics into the visualization research field is something that is not desired by the GRSI because it wants to focus on computer graphics venues (as indicated by its name), then maybe the visualization community could consider to start an own initiative that could collaborate with the GRSI---such that papers in already covered venues such as TVCG, C\&G, and CGF could apply to either group and, if certified, could be cross-listed in both initiatives.\footnote{Maybe a collaboration with initiatives in related fields as RepliCHI in HCI \cite{Wilson:2012:RCF,Wilson:2013:RW,Wilson:2014:RW2} would also be an option---even if RepliCHI (like the EuroRV\textsuperscript{3} workshop) no longer seems to be running.}\textsuperscript{,}\footnote{An alternative to be considered could be the Papers with Code initiative in machine learning at \href{https://paperswithcode.com/}{\texttt{paperswithcode\discretionary{}{.}{.}com}}.}

Finally, there is the potential \textbf{issue that some work is inherently difficult to reproduce}, such as research projects that require special dedicated hardware setups to run (\eg, VR or AR headsets, large wall setups, etc.) that the GRSI is not prepared to test for. I have encountered this very problem myself in the past, and there is no easy solution. Yet often it is possible to circumvent the issue by applying for a GRS for some other aspect of the work, such as the reproduction of the statistical analysis of a user experiment. But maybe the GRSI could consider asking its reviewers to state if they have access to some \emph{standard} special hardware devices such as certain VR or AR headsets, and then to assign such applications to those reviewers who have access to the mentioned setups.

What is interesting to observe is that some authors who have explicitly advocated for reproducibility via the GRSI in the past so far at least have not received a GRS for their papers. I reached out to some of them to ask about the reasons for this lack of papers with GRS. In their responses some noted that they do not publish in the visualization field anymore but in other computer science fields with their own (and partially much lon\-ger-stan\-ding) forms of reproducibility initiatives, so a GRSI certification simply does not apply. Others said that, for some work, they were not the primary supervisor, which relates to the issue of encouragement and incentives I discussed above---a GRSI application usually does not happen without a primary supervisor strongly encouraging the responsible student to apply for the stamp. Yet other work primarily consisted of user experiments, which would call for an actual \emph{replicability} verification which is currently beyond the capability of the GRSI---even if a full study protocol and the associated materials were provided. What this means is that more papers are actually reproducible or replicable than are certified by the GRSI---which would only be discovered by an initiative as the Code Replicability in Computer Graphics (CRCG)\footnote{\href{https://replicability.graphics/}{\texttt{replicability.graphics}}} project \cite{Bonneel:2020:CRC}, which actively and without an application from the authors checks the code resources provided for published papers.

As for the \emph{reproducibility} verification of the statistical analysis of existing study results, which the GRSI does check, the authors I contacted feel strongly that the stamp awarded by the GRSI would be misleading, as they claim to certify \emph{replicability} but in fact they certify \emph{reproducibility}---as I had discussed in \autoref{sec:terminology}. They also feel that the verification suggests a validity that may not exist because authors could also manipulate their data. Here I personally disagree: To some degree we need to trust authors to report the true data. I also feel that a verification of the \emph{reproducibility} of a statistical analysis of the results of an experiment not only enables others to check the numbers or conduct a similar analysis in the future, it also forces the authors of a paper to script, verify, and document the whole analysis code for their very own benefit: By starting with a reproducible analysis setup they become able to quickly sort out problems and re-run the whole analysis with little effort.

The authors also mentioned that \textbf{commercialization intents and proprietary code}\footnote{I can confirm this issue of proprietary code from my own experience, in which case only executable demos are a possibility \cite{Isenberg:2022:PEP}---which are not accepted by the GRSI as reproducibility artifacts.} may prevent a certification, as do the problems of the already mentioned lack of incentives for paper authors. Finally, they pointed at the previously discussed publicity issue---observing that, as of 2024, neither the IEEE VIS call for papers nor the corresponding page on open practices\footnote{\href{https://ieeevis.org/year/2024/info/open-practices/open-practices}{\textls[-20]{\texttt{ieeevis\discretionary{}{.}{.}org\discretionary{/}{}{/}year\discretionary{/}{}{/}2024\discretionary{/}{}{/}info\discretionary{/}{}{/}open\discretionary{}{-}{-}practices\discretionary{/}{}{/}open\discretionary{}{-}{-}practices}}}} currently mention the GRSI, something that easily could be fixed.

So, ultimately, while we have already achieved quite a bit of progress compared to a few years ago, there is still work ahead of us. We need to work toward reducing the obstacles of making work reproducible and, if possible, even replicable, we need to reduce the confusion associated with the many terms being used, we need to increase the awareness of authors of initiatives such as the GRSI, we need to increase the incentives for people to provide the corresponding resources, and we need to discuss if we can extend the scope of venues that are covered by initiatives such as the GRSI.

Of course, any change that addresses the \emph{replication crisis} in the form of more reproducibility or replicability verification also has \textbf{implications for the GRSI} as the initiative that checks it. More applications mean more work, and thus we would need more reviewers to support the initiative by volunteering their time. So we all need to be willing to accept this additional reviewing duty, on top of the existing \emph{reviewing crisis} in (not only) our field. Finally I note that, maybe, the GRSI could consider renaming itself to \emph{Graphics Reproducibility Stamp Initiative} to address the terminology confusion that I had mentioned before in \autoref{sec:terminology}. 

\section{Reproducible paper writing}
\label{sec:analysis-reproducibility}

A paper about research reproducibility would not be authentic without being reproducible itself. I thus ensured that this is the case, not only for the data analysis I describe and the plots I produce, but also for the whole paper itself. I provide all materials, the script and the sources for the paper itself, as well as all external data that I used for my analysis. I normally pull the actually analyzed GRSI data dynamically from the web, and in this case minor manual updates (the country data) are needed for potentially newly added papers to the GRSI website. By default, however, the script runs based on the data status at publication time, but with a small reconfiguration of the script updated data can be used as well. The analysis itself is fully scripted (in a Python script), and produces all images shown in the paper (and more). Some of the numeric analysis results are also collected by the script and then written into \LaTeX{} macros in dedicated files, which are then pulled into the \LaTeX{} paper. So any future \LaTeX{} compilation then uses the updated data, partially also with the \texttt{pgf} package which facilitates in-document calculations---even with the previously mentioned script-generated macros as input data. This process, b.t.w., not only facilitates the reproducibility of the work but also makes the paper writing itself much easier---any change in, correction of, or update of the analysis only needs to be done in or via the analysis script. The paper itself then automatically updates on the next compilation---seemingly by magic---, which means this approach is essentially an equivalent of a Jupyter Notebook for \LaTeX{} document authoring.

To thus be able to reproduce my work, I encourage the reader to get the project from the linked repository (details below), and then to first run the script to do the data analysis and to produce the plots, before compiling the \LaTeX{} sources to reproduce the paper. If the script was instructed to get new data from the web (in this case the country information for the newly added papers would need to be manually added as documented in the \texttt{readme.md}), the paper can be updated to the most recent data without any problems. Naturally, there is a limit to this reproduction process---after a while (likely months to 1--2 years) more authors will have been awarded a GRS and the situation, as I describe it, will have changed, hopefully for the better, and then the discussion in the paper itself will no longer be correct. Moreover, the structure of the website that I query or the used APIs of the digital libraries may also be different then, which could break the processing in the Python script I wrote.

\section*{Data sources}
\label{sec:data_sources}

The data I used in this paper comes from the VisPubData dataset \cite{Isenberg:2017:VMC} as well as from down\-loads\discretionary{/}{}{/}ex\-tracts from the IEEE Xplore, ACM, Elsevier ScienceDirect, Wiley, Eurographics, and Crossref digital libraries; the GRS data comes from \href{https://www.replicabilitystamp.org/}{\texttt{replicabilitystamp\discretionary{}{.}{.}org}}; the data is (and resulting plots are) current as of \GrsiDataCurrentAsOf{}. I also note that the results of the analysis I described in this paper, specifically the fact that a given (IEEE VIS) paper has received a GRS, have already been added to the most recent 2023 edition of the VisPubData dataset \cite{Isenberg:2017:VMC} and will continue to be updated there.

\section*{Supplemental material pointers}
\label{sec:supplemental_materials}

The repository to reproduce the presented results as well as this paper (as discussed in \autoref{sec:analysis-reproducibility}) can be found at \href{https://github.com/tobiasisenberg/Visualization-Reproducibility}{\texttt{github\discretionary{}{.}{.}com\discretionary{/}{}{/}tobiasisenberg\discretionary{/}{}{/}Visualization\discretionary{}{-}{-}Reproducibility}}. A copy of the paper itself including its appendix as well as the figures from the paper can be found at \href{https://osf.io/\osfid/}{\texttt{osf.io/\osfid}}.

\section*{Figure credits and copyright}
\label{sec:figure_credits}

I as the author of this paper state that all figures in this paper are my own as well as are and remain under my own personal copyright, with the permission to be used here. I also make them available under the \href{https://creativecommons.org/licenses/by/4.0/}{Creative Commons At\-tri\-bu\-tion 4.0 International (\ccLogo\,\ccAttribution\ \mbox{CC BY 4.0})} license and share them at \href{https://osf.io/\osfid/}{\texttt{osf.io/\osfid}}.

\acknowledgments{Primarily I wish to thank the Graphics Replicability Stamp Initiative and, in particular, their reviewers for providing their invaluable free service to the computer graphics and visualization communities. In addition, I thank Stefanie Behnke for providing the EuroVis publication data extract from the CGF database, Ross Maciejewski for the list of planned IEEE VIS 2024 TVCG journal presentations, Lane Harrison for the list of planned IEEE VIS 2024 CG\&A journal presentations, the 2024 VIS Papers Chairs for the list of accepted papers, as well as Petra Isenberg and Natkamon Tovanich for their continued work of keeping the VisPubData \cite{Isenberg:2017:VMC} database up to date and error-free. Finally, thanks to Lonni Besan\c{c}on, Martin Skrodzki, and Petra Isenberg for feedback on this manuscript, to the authors whose replies I cite anonymously in \autoref{sec:discussion} for their answers, as well as Cody Dunne for help with polishing the documentation of the \href{https://github.com/tobiasisenberg/Visualization-Reproducibility}{GitHub repository} and with fixing some issues with the code for specific Python environments.}

\bibliographystyle{abbrv-doi-hyperref}

\bibliography{abbreviations,template}

\begin{thebibliography}{10}

\bibitem{Besancon:2021:OSS}
L.~Besan\c{c}on, N.~Peiffer-Smadja, C.~Segalas, H.~Jiang, P.~Masuzzo, C.~Smout,
  E.~Billy, M.~Deforet, and C.~Leyrat.
\newblock Open science saves lives: Lessons from the {COVID}-19 pandemic.
\newblock {\em BMC Med Res Methodol}, 21,  art. no. 117,  18 pages, 2021.
  \href{https://doi.org/10/gkdzr6}
{doi: {{%
10\discretionary{/}{%
}{/}gkdzr6}}}


\bibitem{Boisvert:2016:IR}
R.~F. Boisvert.
\newblock Incentivizing reproducibility.
\newblock {\em Commun ACM}, 59(10):5, 2016. \href{https://doi.org/10/gc5pts}
{doi: {{%
10\discretionary{/}{%
}{/}gc5pts}}}


\bibitem{Bonneel:2020:CRC}
N.~Bonneel, D.~Coeurjolly, J.~Digne, and N.~Mellado.
\newblock Code replicability in computer graphics.
\newblock {\em ACM Trans Graph}, 39(4),  art. no. 93,  8 pages, 2020.
  \href{https://doi.org/10/gg8xfh}
{doi: {{%
10\discretionary{/}{%
}{/}gg8xfh}}}


\bibitem{Chopra:2023:PGG}
S.~Chopra, L.~Labache, E.~Dhamala, E.~R. Orchard, and A.~Holmes.
\newblock A practical guide for generating reproducible and programmatic
  neuroimaging visualizations.
\newblock {\em Aperture Neuro}, 3,  art. no. 001c.85104,  20 pages, 2023.
  \href{https://doi.org/10/gtw4sn}
{doi: {{%
10\discretionary{/}{%
}{/}gtw4sn}}}


\bibitem{Claerbout:1992:EDG}
J.~F. Claerbout and M.~Karrenbach.
\newblock Electronic documents give reproducible research a new meaning.
\newblock In {\em SEG Expanded Abstracts}, pp. 601--604. Society of Exploration
  Geophysicists, Houston, 1992. \href{https://doi.org/10/b6t7wj}
{doi: {{%
10\discretionary{/}{%
}{/}b6t7wj}}}


\bibitem{Cockburn:2020:TRC}
A.~Cockburn, P.~Dragicevic, L.~Besan\c{c}on, and C.~Gutwin.
\newblock Threats of a replication crisis in empirical computer science.
\newblock {\em Commun ACM}, 63(8):70--79, 2020.
  \href{https://doi.org/10/gjbnx4}
{doi: {{%
10\discretionary{/}{%
}{/}gjbnx4}}}


\bibitem{Crumwell:2023:WBC}
S.~Cr{\"u}well, D.~Apthorp, B.~J. Baker, L.~Colling, M.~Elson, S.~J. Geiger,
  S.~Lobentanzer, J.~Mon{\'e}ger, A.~Patterson, D.~S. Schwarzkopf, M.~Zaneva,
  and N.~J.~L. Brown.
\newblock What's in a badge? {A} computational reproducibility investigation of
  the open data badge policy in one issue of psychological science.
\newblock {\em Psychol Sci}, 34(4):512--522, 2023.
  \href{https://doi.org/10/grzd4n}
{doi: {{%
10\discretionary{/}{%
}{/}grzd4n}}}


\bibitem{Cushing:2018:SVR}
J.~B. Cushing, D.~Lach, C.~Zanocco, and J.~Halama.
\newblock Scientific visualization and reproducibility for ``open''
  environmental science.
\newblock In {\em Proc.\ Big Data}, pp. 3211--3216. IEEE CS, Los Alamitos,
  2018. \href{https://doi.org/10/gtw4zm}
{doi: {{%
10\discretionary{/}{%
}{/}gtw4zm}}}


\bibitem{DeFloriani:2018:MEC}
L.~De~Floriani.
\newblock State of the journal.
\newblock {\em IEEE Trans Vis Comput Graph}, 24(2):1036--1037, 2018.
  \href{https://doi.org/10/gtzdcx}
{doi: {{%
10\discretionary{/}{%
}{/}gtzdcx}}}


\bibitem{Fekete:2020:ERV}
J.-D. Fekete and J.~Freire.
\newblock Exploring reproducibility in visualization.
\newblock {\em IEEE Comput Graph Appl}, 40(5):108--119, 2020.
  \href{https://doi.org/10/ghd59m}
{doi: {{%
10\discretionary{/}{%
}{/}ghd59m}}}


\bibitem{Franke:2023:TRV}
M.~Franke, G.~Reina, and S.~Koch.
\newblock Toward reproducible visual analysis results.
\newblock In {\em Proc.\ PacificVis}, pp. 102--106. IEEE CS, Los Alamitos,
  2023. \href{https://doi.org/10/gtw42q}
{doi: {{%
10\discretionary{/}{%
}{/}gtw42q}}}


\bibitem{Garkov:2022:RDC}
D.~Garkov, C.~Müller, M.~Braun, D.~Weiskopf, and F.~Schreiber.
\newblock Research data curation in visualization: Position paper.
\newblock In {\em Proc.\ BELIV}, pp. 56--65. IEEE CS, Los Alamitos, 2022.
  \href{https://doi.org/10/gttm5m}
{doi: {{%
10\discretionary{/}{%
}{/}gttm5m}}}


\bibitem{Glencross:2022:GBP}
M.~Glencross, M.~Attene, D.~Panozzo, M.~C. Lin, and A.~Gomes.
\newblock {GRSI} best paper award.
\newblock {\em Comput Graph}, 108:A6, 2022. \href{https://doi.org/10/nfqg}
{doi: {{%
10\discretionary{/}{%
}{/}nfqg}}}


\bibitem{Glencross:2021:GBP}
M.~Glencross, D.~Panozzou, and J.~Jorge.
\newblock {GRSI} best paper award.
\newblock {\em Comput Graph}, 99:A5--A6, 2021. \href{https://doi.org/10/nfqf}
{doi: {{%
10\discretionary{/}{%
}{/}nfqf}}}


\bibitem{Haroz:2018:OPV}
S.~Haroz.
\newblock Open practices in visualization research: Opinion paper.
\newblock In {\em Proc.\ BELIV}, pp. 46--52. IEEE CS, Los Alamitos, 2018.
  \href{https://doi.org/10/gtw4sp}
{doi: {{%
10\discretionary{/}{%
}{/}gtw4sp}}}


\bibitem{Isenberg:2017:VMC}
P.~Isenberg, F.~Heimerl, S.~Koch, T.~Isenberg, P.~Xu, C.~D. Stolper,
  M.~Sedlmair, J.~Chen, T.~M{\"o}ller, and J.~Stasko.
\newblock vispubdata.org: A metadata collection about {IEEE} visualization
  ({VIS}) publications.
\newblock {\em IEEE Trans Vis Comput Graph}, 23(9):2199--2206, 2017.
  \href{https://doi.org/10/ggwwrv}
{doi: {{%
10\discretionary{/}{%
}{/}ggwwrv}}}


\bibitem{Isenberg:2022:PEP}
T.~Isenberg.
\newblock Personal experiences of providing and using research prototypes.
\newblock In {\em Proc.\ VisGap}, pp. 17--22. EG, Goslar, 2022.
  \href{https://doi.org/10/gr2d83}
{doi: {{%
10\discretionary{/}{%
}{/}gr2d83}}}


\bibitem{Jansen:2024:MWB}
Y.~Jansen, J.~B. Vornhagen, O.~Iarygina, K.~S. Niksirat, L.~Besan\c{c}on,
  P.~Dragicevic, J.~Gori, and C.~Wacharamanotham.
\newblock The many ways of being transparent in human-computer interaction
  research.
\newblock {OSF} preprint, 2024. \href{https://doi.org/10/gt3f2v}
{doi: {{%
10\discretionary{/}{%
}{/}gt3f2v}}}


\bibitem{Jorge:2018:EN}
J.~Jorge.
\newblock Editorial note.
\newblock {\em Computers \& Graphics}, 74:A1--A2, 2018.
  \href{https://doi.org/10/gtzdcz}
{doi: {{%
10\discretionary{/}{%
}{/}gtzdcz}}}


\bibitem{Jorge:2020:NEC}
J.~Jorge.
\newblock A note from the editor in chief.
\newblock {\em Computers \& Graphics}, 88:A1--A3, 2020.
  \href{https://doi.org/10/gtzdc3}
{doi: {{%
10\discretionary{/}{%
}{/}gtzdc3}}}


\bibitem{Kosara:2016:EBS}
R.~Kosara.
\newblock An empire built on sand: Reexamining what we think we know about
  visualization.
\newblock In {\em Proc.\ BELIV}, pp. 162--168. ACM, New York, 2016.
  \href{https://doi.org/10/gfz5kr}
{doi: {{%
10\discretionary{/}{%
}{/}gfz5kr}}}


\bibitem{Kosara:2018:SRC}
R.~Kosara and S.~Haroz.
\newblock Skipping the replication crisis in visualization: Threats to study
  validity and how to address them: Position paper.
\newblock In {\em Proc.\ BELIV}, pp. 102--107. IEEE CS, Los Alamitos, 2018.
  \href{https://doi.org/10/gtw4sq}
{doi: {{%
10\discretionary{/}{%
}{/}gtw4sq}}}


\bibitem{Plesser:2018:RRB}
H.~E. Plesser.
\newblock Reproducibility vs. replicability: A brief history of a confused
  terminology.
\newblock {\em Front Neuroinf}, 11,  art. no. 76,  4 pages, 2018.
  \href{https://doi.org/10/gc5ptr}
{doi: {{%
10\discretionary{/}{%
}{/}gc5ptr}}}


\bibitem{Quadri:2019:YCP}
G.~J. Quadri and P.~Rosen.
\newblock You can’t publish replication studies (and how to anyways).
\newblock arXiv preprint 1908.08893, 2019. \href{https://doi.org/10/gtxgh9}
{doi: {{%
10\discretionary{/}{%
}{/}gtxgh9}}}


\bibitem{Reina:2023:CID}
G.~Reina.
\newblock Can image data facilitate reproducibility of graphics and
  visualizations? {T}oward a trusted scientific practice.
\newblock {\em IEEE Comput Graph Appl}, 43(2):89--99, 2023.
  \href{https://doi.org/10/gtw4nt}
{doi: {{%
10\discretionary{/}{%
}{/}gtw4nt}}}


\bibitem{Rougier​:2017:SCS}
N.~P. Rougier, K.~Hinsen, F.~Alexandre, T.~Arildsen, L.~A. Barba, F.~C.~Y.
  Benureau, C.~T. Brown, P.~de~Buyl, O.~Caglayan, A.~P. Davison, M.-A. Delsuc,
  G.~Detorakis, A.~K. Diem, D.~Drix, P.~Enel, B.~Girard, O.~Guest, M.~G. Hall,
  R.~N. Henriques, X.~Hinaut, K.~S. Jaron, M.~Khamassi, A.~Klein, T.~Manninen,
  P.~Marchesi, D.~McGlinn, C.~Metzner, O.~Petchey, H.~E. Plesser, T.~Poisot,
  K.~Ram, Y.~Ram, E.~Roesch, C.~Rossant, V.~Rostami, A.~Shifman, J.~Stachelek,
  M.~Stimberg, F.~Stollmeier, F.~Vaggi, G.~Viejo, J.~Vitay, A.~E. Vostinar,
  R.~Yurchak, and T.~Zito.
\newblock Sustainable computational science: The {ReScience} initiative.
\newblock {\em PeerJ Comput Sci}, 3,  art. no. e142,  17 pages, 2017.
  \href{https://doi.org/10/gcx5kf}
{doi: {{%
10\discretionary{/}{%
}{/}gcx5kf}}}


\bibitem{Sukumar:2018:TDU}
P.~T. Sukumar and R.~Metoyer.
\newblock Towards designing unbiased replication studies in information
  visualization.
\newblock In {\em Proc.\ BELIV}, pp. 93--101. IEEE CS, Los Alamitos, 2018.
  \href{https://doi.org/10/gtw4sr}
{doi: {{%
10\discretionary{/}{%
}{/}gtw4sr}}}


\bibitem{Syeda:2024:VRT}
U.~H. Syeda, L.~South, J.~Raynor, L.~Panavas, D.~Saffo, T.~Morriss, C.~Dunne,
  and M.~A. Borkin.
\newblock Vis repligogy: Towards a culture of facilitating replication studies
  in visualization pedagogy and research.
\newblock In {\em EuroVis Education Papers},  art. no. eved.20241054,  9 pages.
  EG, Goslar, 2024. \href{https://doi.org/10/gtxbsb}
{doi: {{%
10\discretionary{/}{%
}{/}gtxbsb}}}


\bibitem{Valdez:2018:RRR}
A.~C. Valdez, A.~K. Schaar, J.~R. Hildebrandt, and M.~Ziefle.
\newblock Requirements for reproducibility of research in situational and
  spatio-temporal visualization: Position paper.
\newblock In {\em Proc.\ BELIV}, pp. 53--59. IEEE CS, Los Alamitos, 2018.
  \href{https://doi.org/10/gtw4ss}
{doi: {{%
10\discretionary{/}{%
}{/}gtw4ss}}}


\bibitem{Wilson:2014:RW2}
M.~L. Wilson, E.~H. Chi, S.~Reeves, and D.~Coyle.
\newblock {RepliCHI}: The workshop {II}.
\newblock In {\em CHI Extended Abstracts}, pp. 33--36. ACM, New York, 2014.
  \href{https://doi.org/10/gt28zz}
{doi: {{%
10\discretionary{/}{%
}{/}gt28zz}}}


\bibitem{Wilson:2012:RCF}
M.~L. Wilson, W.~Mackay, E.~H. Chi, M.~Bernstein, and J.~Nichols.
\newblock {RepliCHI} {SIG}: From a panel to a new submission venue for
  replication.
\newblock In {\em CHI Extended Abstracts}, pp. 1185--1188. ACM, New York, 2012.
  \href{https://doi.org/10/m59w}
{doi: {{%
10\discretionary{/}{%
}{/}m59w}}}


\bibitem{Wilson:2013:RW}
M.~L. Wilson, P.~Resnick, D.~Coyle, and E.~H. Chi.
\newblock {RepliCHI}: The workshop.
\newblock In {\em CHI Extended Abstracts}, pp. 3159--3162. ACM, New York, 2013.
  \href{https://doi.org/10/gt28zx}
{doi: {{%
10\discretionary{/}{%
}{/}gt28zx}}}


\end{thebibliography}


\appendix 

\section{Keywords in visualization classification}
\label{app:keywords}

If the following keywords were present in the paper \emph{title} and the paper had not been presented at a visualization conference, then I still classified the paper as talking about visualization topics:
\begin{itemize}[parsep=0pt]
\item \emph{visualization}
\item \emph{visualisation}
\item \emph{visualizing}
\item \emph{visualising}
\item \emph{visual} \texttt{AND} \emph{analytics}
\item \emph{visual} \texttt{AND} \emph{analysis}
\item \emph{visual representation}
\item \emph{data exploration}
\item \emph{visual exploration}
\item \emph{graph drawing}
\item \emph{parallel coordinate}s
\item \emph{scatterplot}
\item \emph{choropleth}
\item \emph{cartogram}
\item \emph{star glyph}
\item \emph{glyph design}
\item \emph{line graph}
\item \emph{streamgraph}
\item \emph{focus+context}
\item \emph{t-sne}
\item \emph{high-dimensional data}
\end{itemize}

This keyword-based classification found the following \GrsiVisByKeywordPapersCount{} papers (out of the \GrsiTotalVisPapers{} visualization papers in my analysis; \ie, \percentageRounded{\GrsiVisByKeywordPapersCount}{\GrsiTotalVisPapers}\%) that I had not previously identified as having been presented by their authors at a clearly visualization-themed conference:
\begin{itemize}[parsep=0pt]
\item \href{https://doi.org/10.1016/j.cag.2024.104058}{doi: 10.1016/j.cag.2024.104058}
\item \href{https://doi.org/10.1109/tvcg.2024.3418653}{doi: 10.1109/tvcg.2024.3418653}
\item \href{https://doi.org/10.1109/tvcg.2023.3237768}{doi: 10.1109/tvcg.2023.3237768}
\item \href{https://doi.org/10.1109/tvcg.2023.3345532}{doi: 10.1109/tvcg.2023.3345532}
\item \href{https://doi.org/10.1016/j.cag.2022.10.008}{doi: 10.1016/j.cag.2022.10.008}
\item \href{https://doi.org/10.1111/cgf.14487}{doi: 10.1111/cgf.14487}
\item \href{https://doi.org/10.1111/cgf.14615}{doi: 10.1111/cgf.14615}
\item \href{https://doi.org/10.1109/tvcg.2022.3151227}{doi: 10.1109/tvcg.2022.3151227}
\item \href{https://doi.org/10.1109/tvcg.2022.3155564}{doi: 10.1109/tvcg.2022.3155564}
\item \href{https://doi.org/10.1109/tvcg.2021.3122388}{doi: 10.1109/tvcg.2021.3122388}

\end{itemize}

\section{Reasons for manual visualization classification}
\label{app:manual-reasons}

The presentation-based or keyword-based classification did not find some papers which I still consider to be visualization papers. Here are these \GrsiVisManuallyMarkedPapersCount{} papers (out of the \GrsiTotalVisPapers{} visualization papers; \ie, \percentageRounded{\GrsiVisManuallyMarkedPapersCount}{\GrsiTotalVisPapers}\%) and the respective reasons I classified them as talking about visualization topics:
\begin{itemize}[parsep=0pt]
\item \href{https://doi.org/10.1109/tvcg.2022.3214821}{doi: 10.1109/tvcg.2022.3214821}: \emph{visualization} used as author keyword
\item \href{https://doi.org/10.1109/tvcg.2021.3101418}{doi: 10.1109/tvcg.2021.3101418}: \emph{visualization} used as author keyword
\item \href{https://doi.org/10.1109/tvcg.2021.3067820}{doi: 10.1109/tvcg.2021.3067820}: \emph{visualization} in the paper abstract
\item \href{https://doi.org/10.1016/j.cag.2024.01.001}{doi: 10.1016/j.cag.2024.01.001}: \emph{visualization} in the paper abstract
\item \href{https://doi.org/10.1109/tvcg.2020.2966702}{doi: 10.1109/tvcg.2020.2966702}: is on flattening of 3D surfaces from data
\item \href{https://doi.org/10.1016/j.cag.2023.06.023}{doi: 10.1016/j.cag.2023.06.023}: talks about molecular channel datasets
\item \href{https://doi.org/10.1145/3528223.3530102}{doi: 10.1145/3528223.3530102}: talks about simulation and visualization of stellar atmospheres
\item \href{https://doi.org/10.1111/cgf.14784}{doi: 10.1111/cgf.14784}: talks about topology, graphs, and scalar fields
\item \href{https://doi.org/10.1111/cgf.13910}{doi: 10.1111/cgf.13910}: talks about point clouds and topology
\end{itemize}
This list shows that I essentially classified only five papers manually, the other four (the top four in the list) use the \emph{visualization} term in either the author keywords or the paper abstract (but which I cannot reliably get from the online publication meta data, so had to partially manually add this to the data and thus made a manual classification).

\section{Additional Plots}
\label{app:plots}

\begin{figure*}
	\centering
	\includegraphics[width=\linewidth]{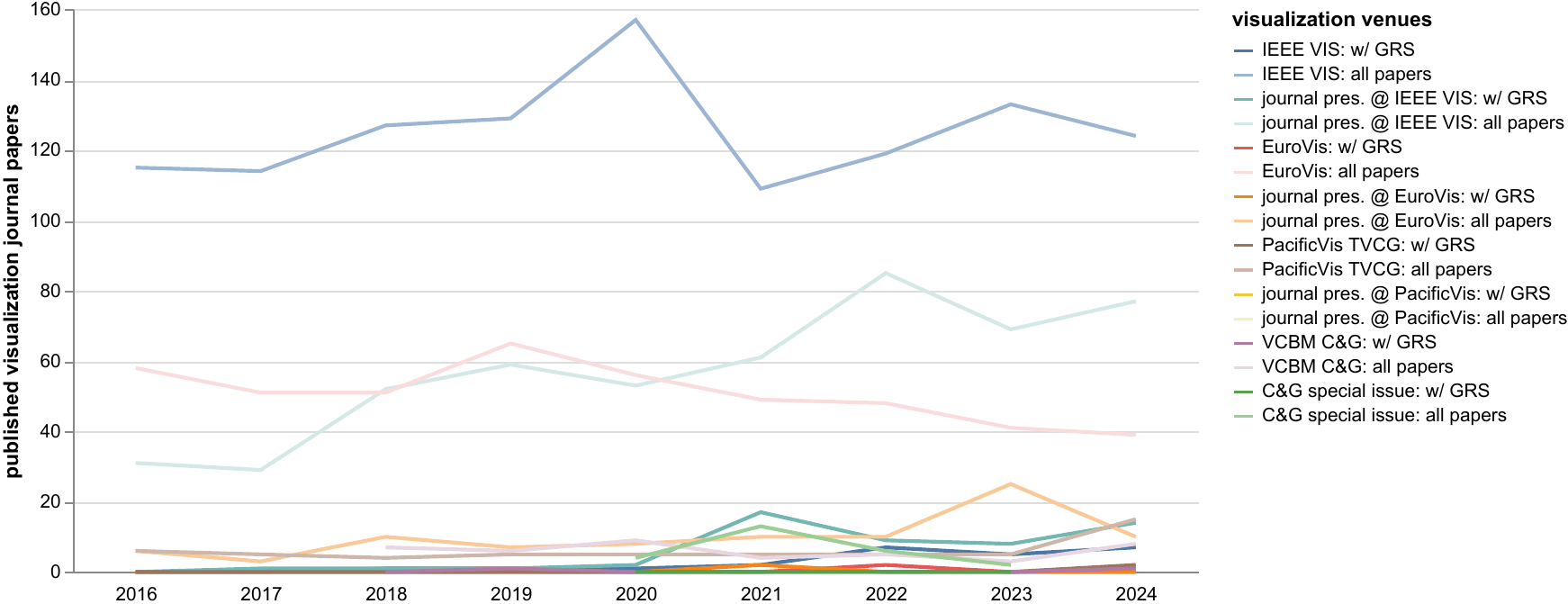}
	\caption{Line graph version of \autoref{fig:teaser} (but without the stacking aspect of \autoref{fig:teaser}, so here I show one line for only the GRSI-awarded papers of the venue plus one line for all papers of a venue): Papers that can clearly be classified as visualization work (visualization conferences and special issues in journals) by \emph{presentation} year (\ie, based on the \emph{conference years}, in which the papers were presented) and their subset with GRS.\textsuperscript{\ref{foot:vis_presentations},\ref{foot:c-and-g-special}}}
	\label{fig:vis-grs-presentations-by-year-linegraph}
\end{figure*}

\begin{figure*}
	\centering
	\includegraphics[width=\linewidth]{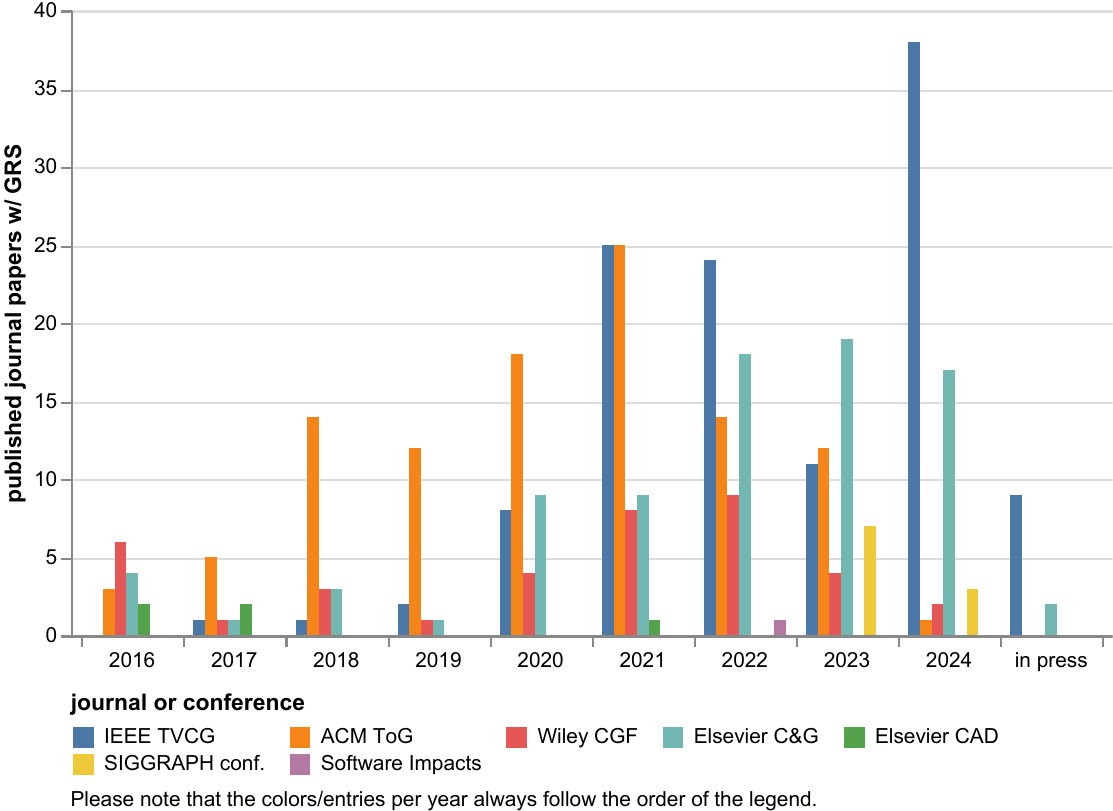}
	\caption{Bar chart version of \autoref{fig:grs-overall}: Overall development of papers with GRS, by publication venues (and their article \emph{publication} years).}
	\label{fig:grs-overall-bars}
\end{figure*}

\begin{figure*}
	\centering
	\includegraphics[height=1.3\columnwidth]{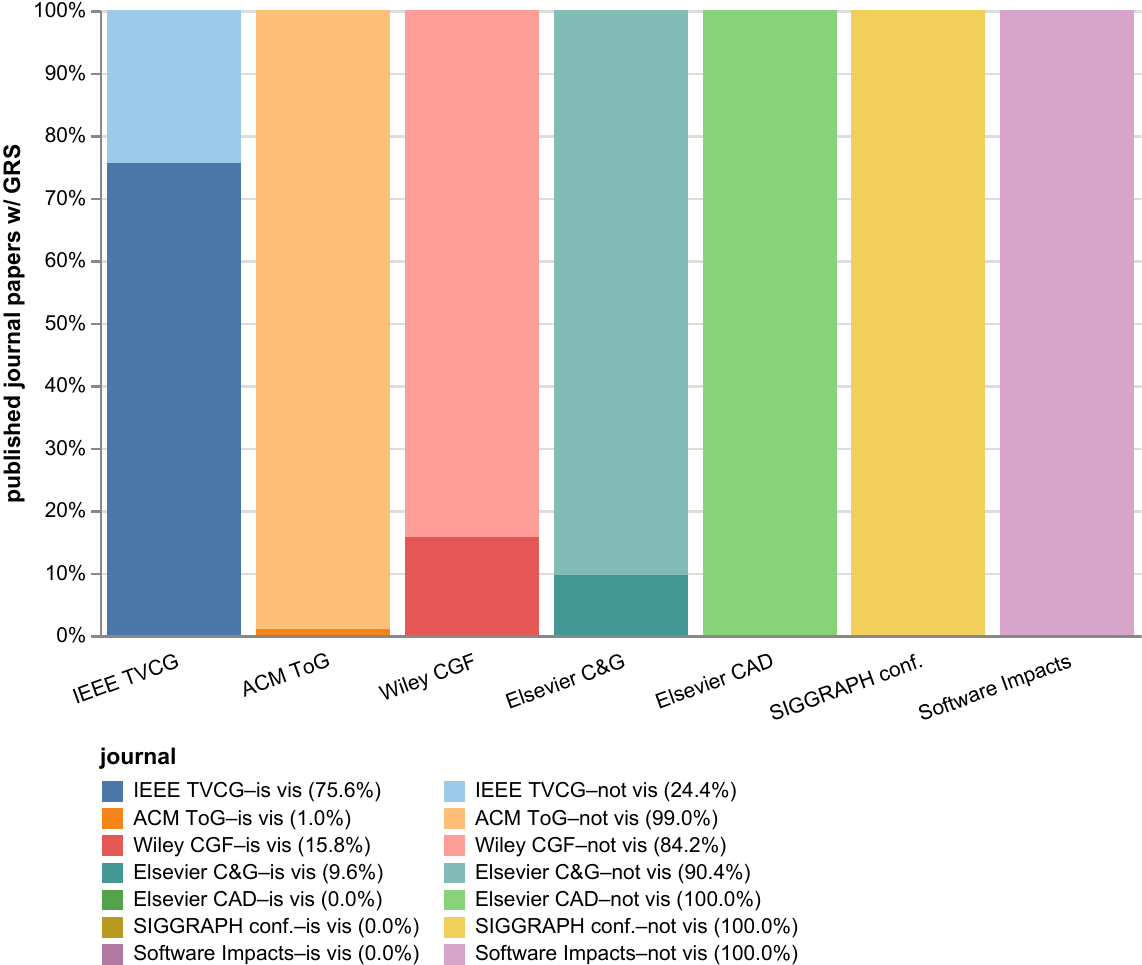}
	\caption{Normalized version of \autoref{fig:grs-split-vis-novis-stackedbar}: GRS papers by publication venue and their respective proportion classified as visualization papers.}
	\label{fig:grs-split-vis-novis-stackedbar-normalized}
\end{figure*}

\begin{figure*}
	\centering
	\includegraphics[height=.75\columnwidth,trim={0 0 190px 0},clip]{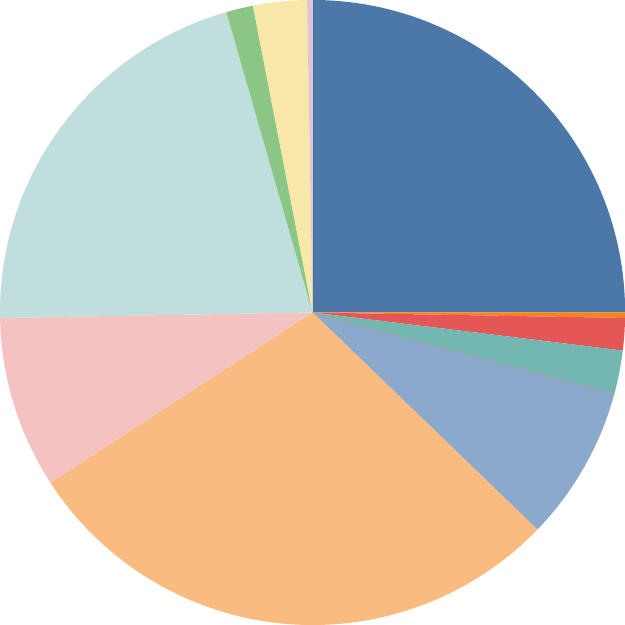}\hfill%
	\includegraphics[height=.75\columnwidth]{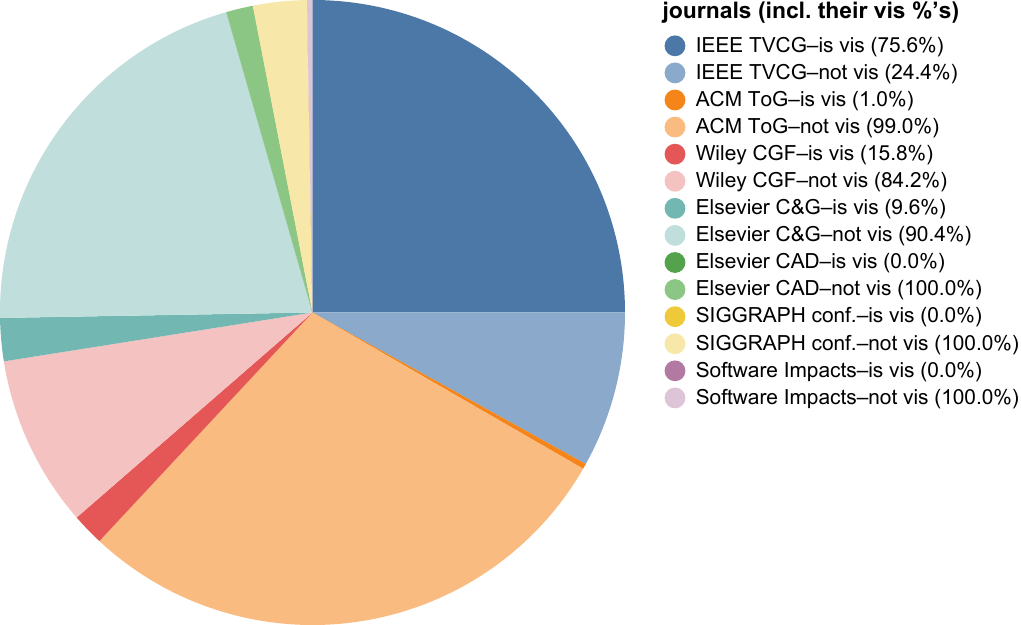}
	\caption{Proportion of GRS papers classified as visualization papers per publication venue. Left: sorted by visualization status, right: sorted by publication venue. This is essentially the same data as in \autoref{fig:grs-split-vis-novis-stackedbar}, only shown as a pie chart to better portray the percentages.}
	\label{fig:grs-split-vis-novis-piechart}
\end{figure*}

\begin{figure*}
	\centering
	\includegraphics[height=1.2\columnwidth]{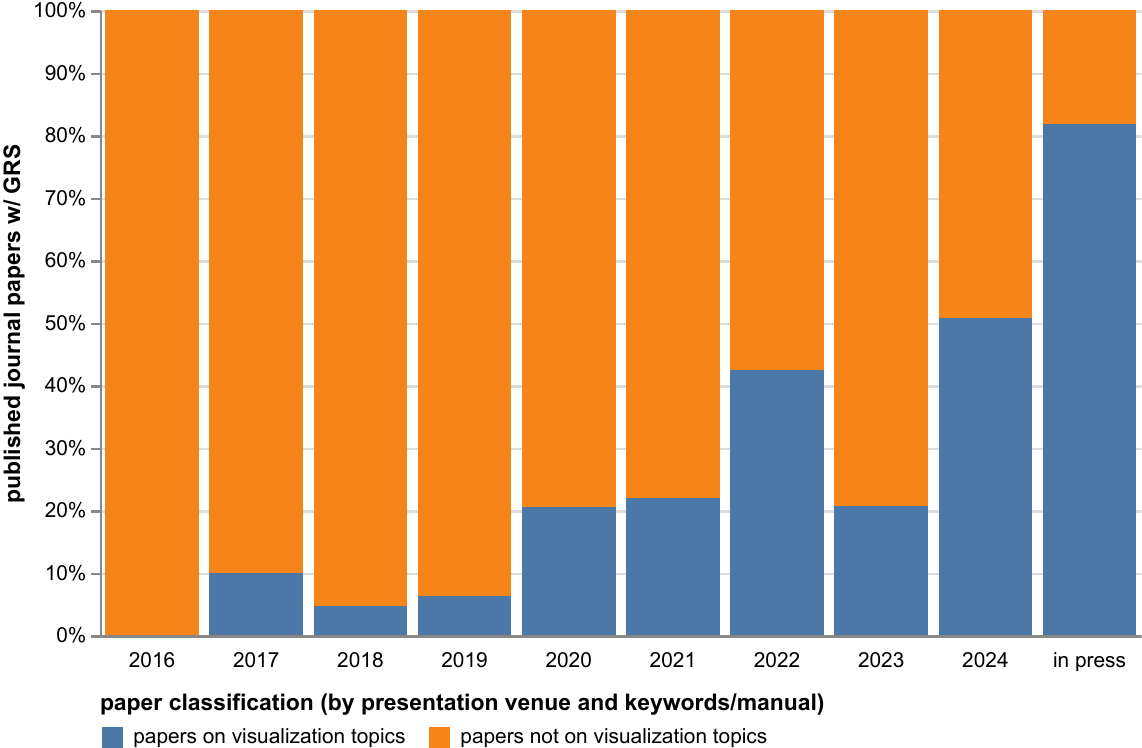}
	\caption{Normalized version of \autoref{fig:grs-split-vis-novis}: Overall proportion of GRS papers classified as visualization papers by article \emph{publication} years.}
	\label{fig:grs-split-vis-novis-normalized}
\end{figure*}

\begin{figure*}
	\centering
	\includegraphics[height=1.3\columnwidth]{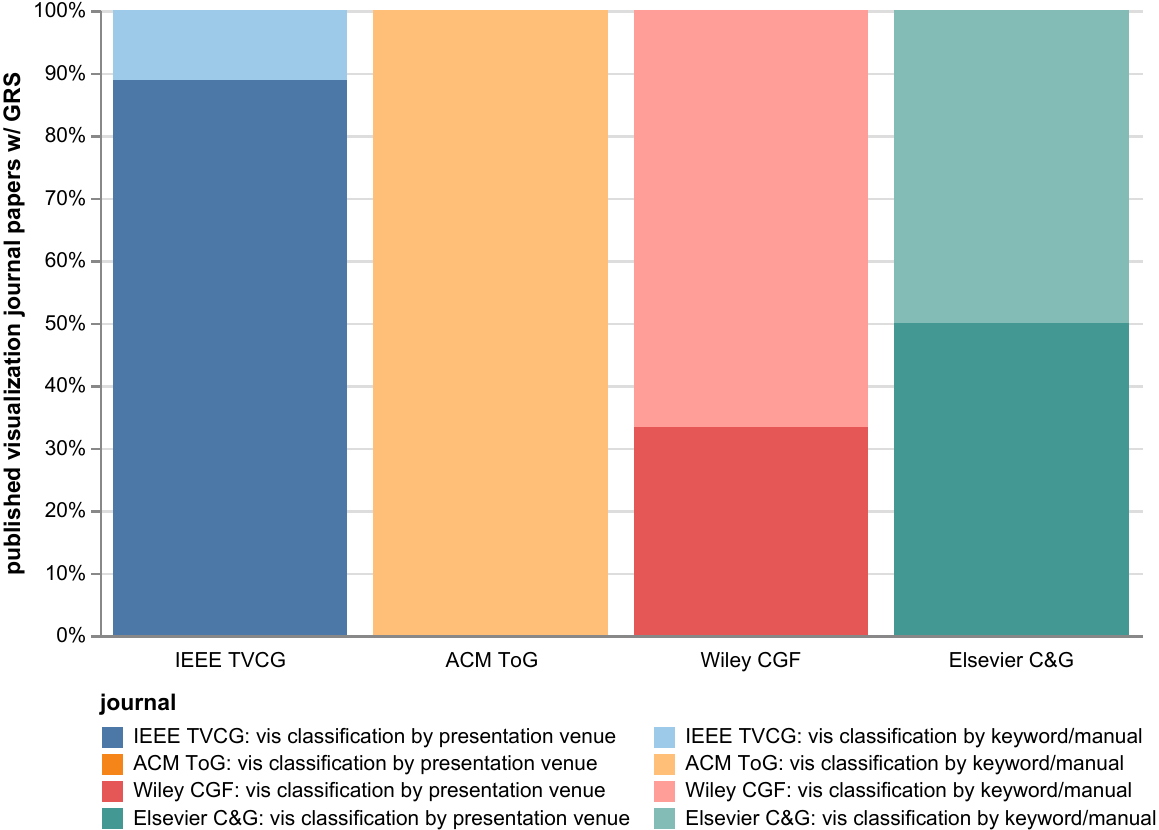}
	\caption{Normalized version of \autoref{fig:vis-grs-journals-aggregated}: Sources of visualization papers overall, split between classification by visualization presentation venue and keyword/manual classification.}
	\label{fig:vis-grs-journals-aggregated-normalized}
\end{figure*}

\begin{figure*}
	\centering
	\includegraphics[height=1.2\columnwidth]{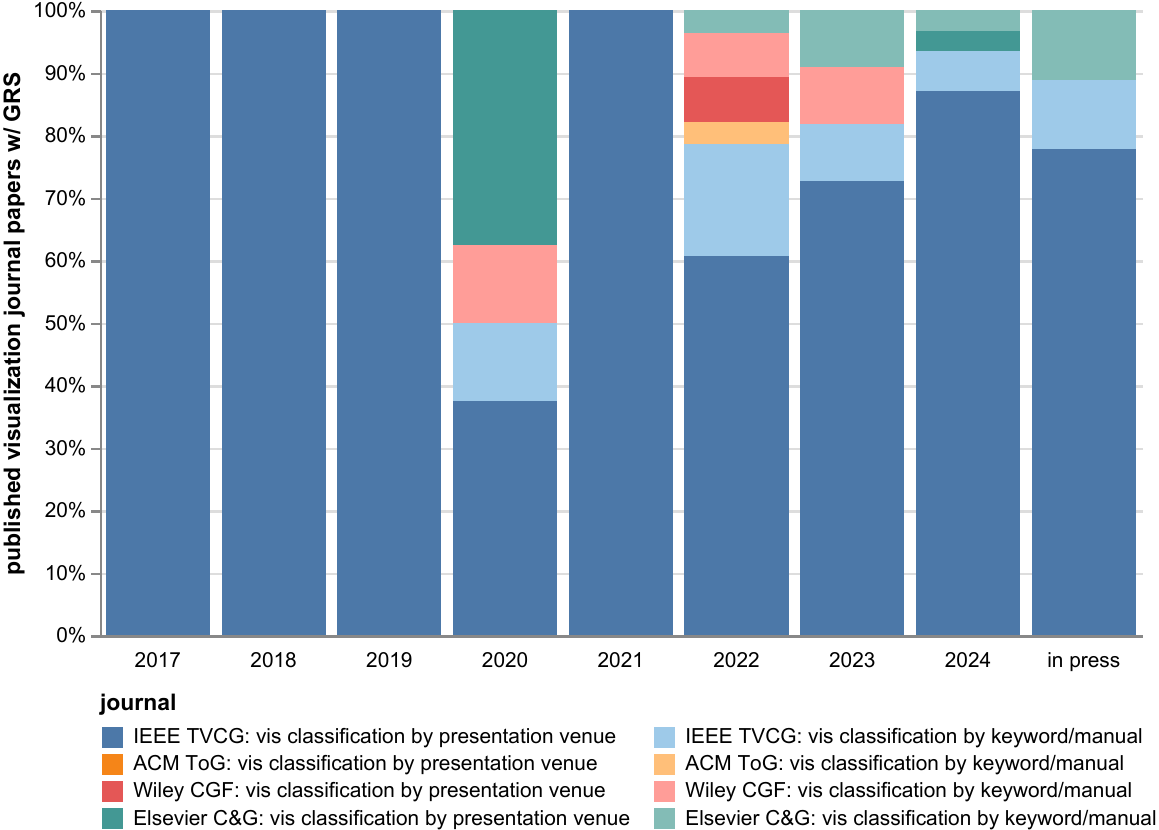}
	\caption{Normalized version of \autoref{fig:vis-grs-journals-by-year}: Sources of visualization papers overall, split between classification by visualization presentation venue and keyword/manual classification, by article \emph{publication} year.}
	\label{fig:vis-grs-journals-by-year-normalized}
\end{figure*}

\begin{figure*}
	\centering
	\includegraphics[height=1.2\columnwidth]{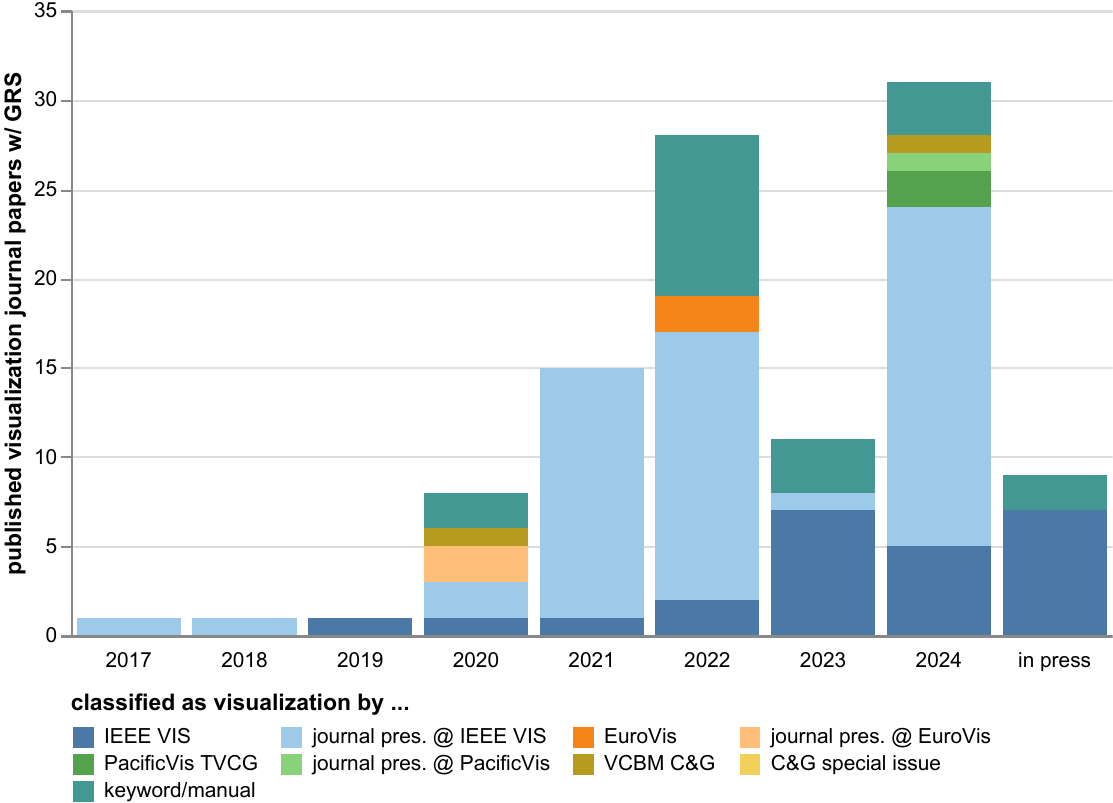}
	\caption{Stacked bar chart version of \autoref{fig:grs-vis-classification}: Different types of classification of papers as visualization work, by article \emph{publication} year (\ie, not by paper \emph{presentation} year).}
	\label{fig:grs-vis-classification-barchart}
\end{figure*}

\begin{figure*}
	\centering
	\includegraphics[height=1.2\columnwidth]{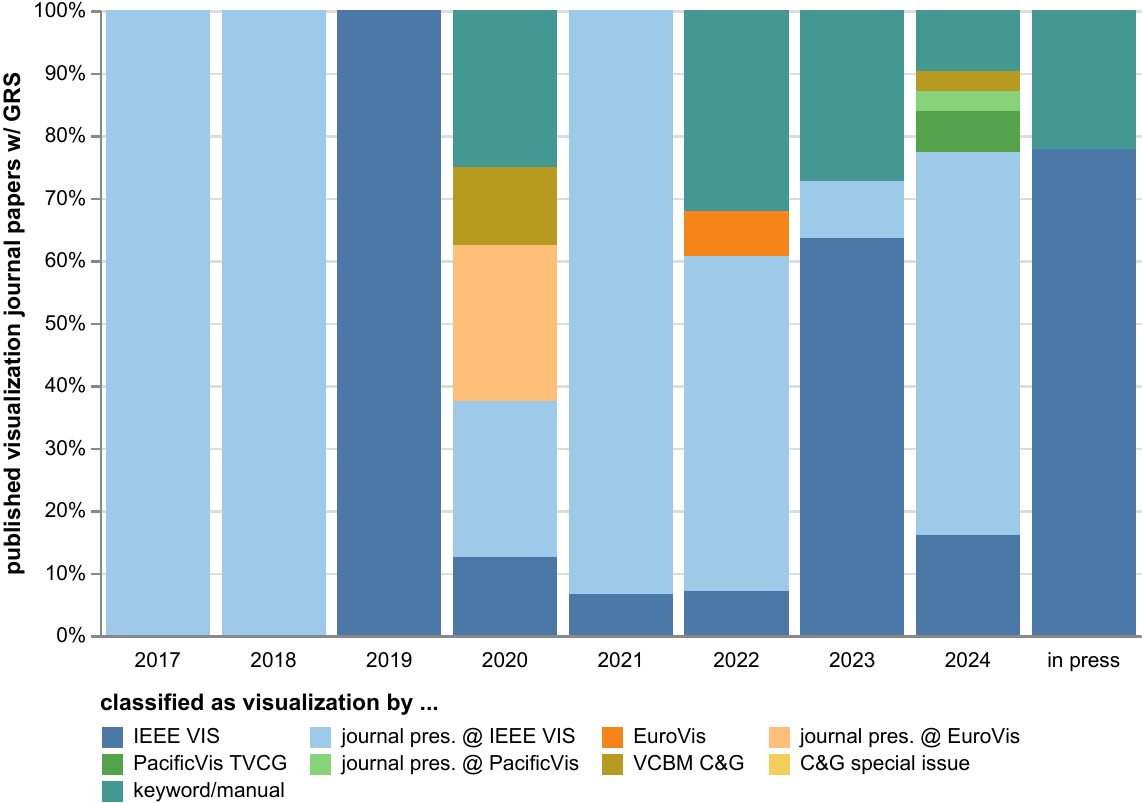}
	\caption{Normalized version of \autoref{fig:grs-vis-classification-barchart}: Different types of classification of papers as visualization work, by article \emph{publication} year (\ie, not by paper \emph{presentation} year).}
	\label{fig:grs-vis-classification-barchart-normalized}
\end{figure*}

\begin{figure*}
	\centering
	\includegraphics[height=.4\linewidth]{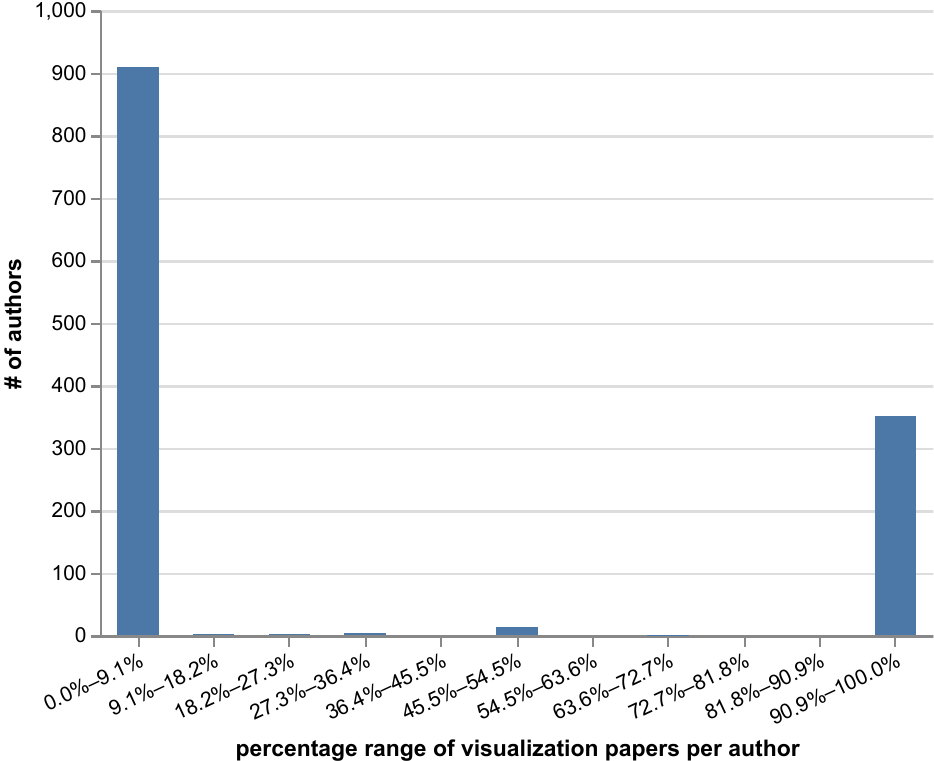}\hfill%
	\includegraphics[height=.4\linewidth]{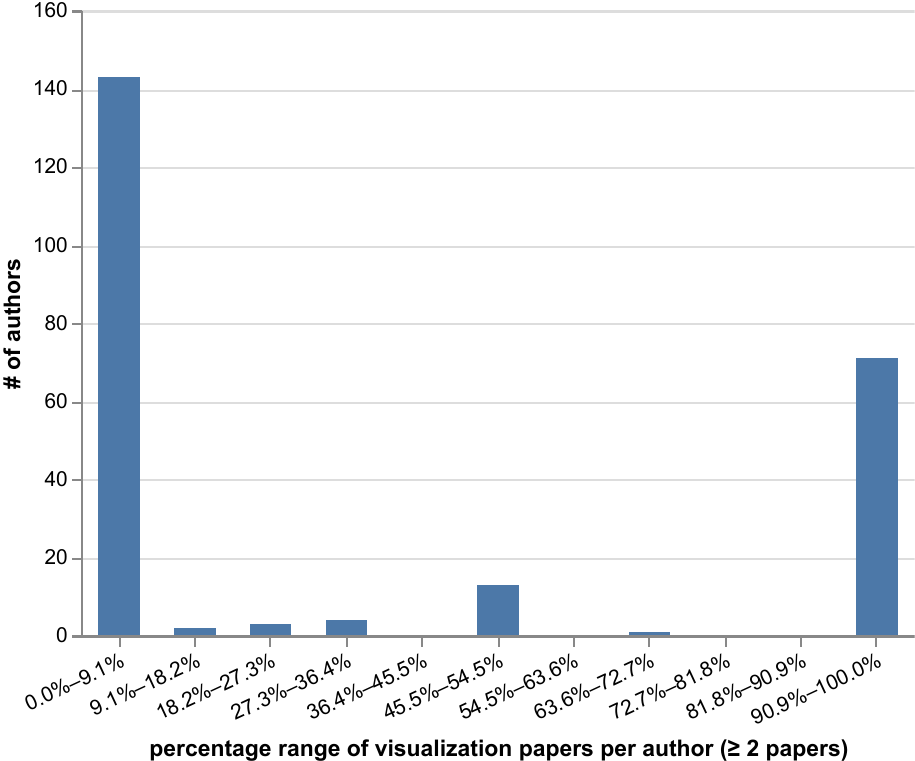}%
	\caption{Same data as in \autoref{fig:histogram-author-vis-percentage} but linearly plotted: Histogram of the percentage of visualization papers per author (left: all authors, right: only authors with $\geq$ 2 papers).}
	\label{fig:histogram-author-vis-percentage-nolog}
\end{figure*}

\begin{figure*}
	\centering
	\includegraphics[height=.53\columnwidth]{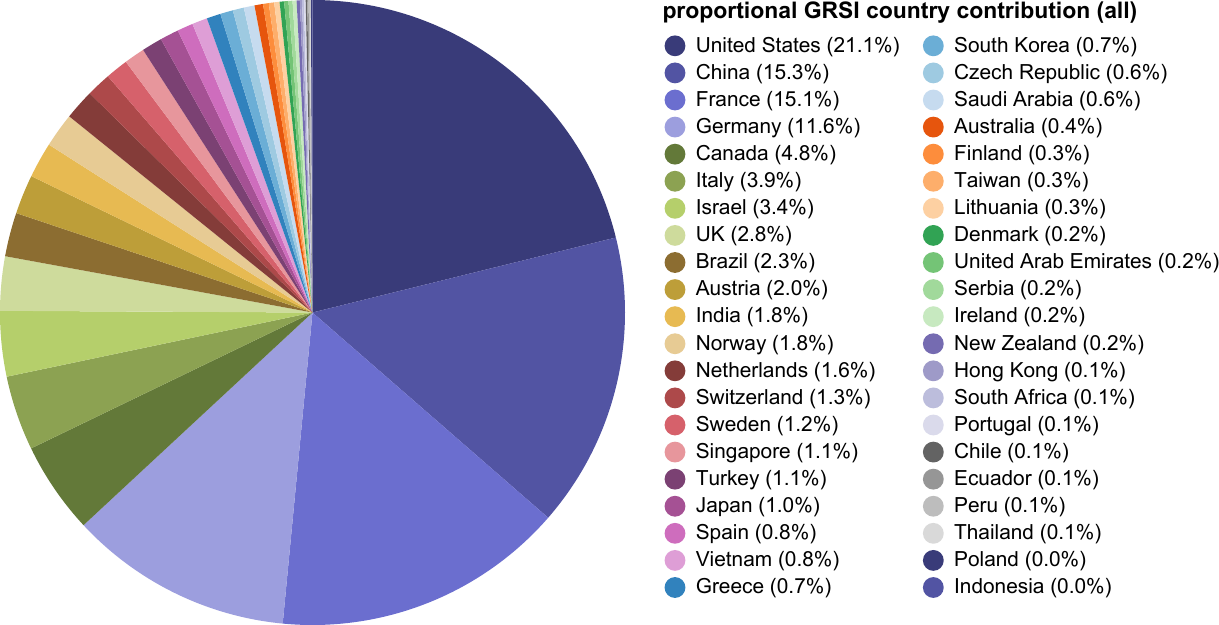}\hfill%
	\includegraphics[height=.53\columnwidth]{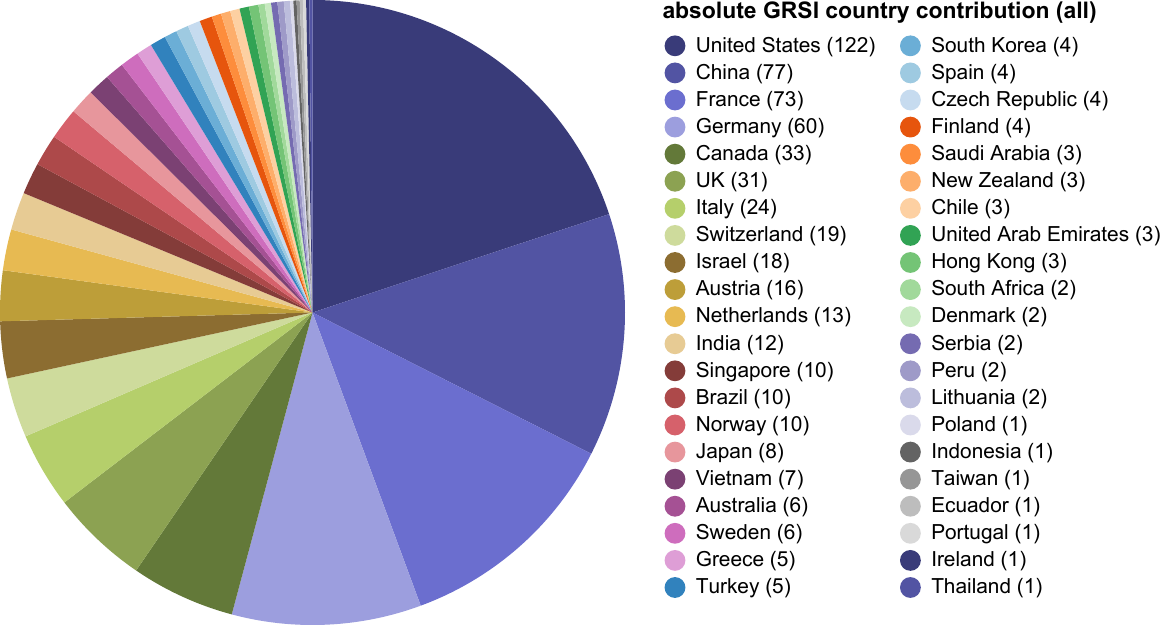}
	\caption{Proportional (left) and absolute (right) country contribution to GRS awards, looking at all paper authors. The latter counts each paper with a co-author from a country for that country (each paper being counted once per country at most), while the former pro-rates the contributions among all co-authors of a papers and also potentially among their individual shared co-affiliations as described in \autoref{sec:visual_analysis}. The left image is a more detailed version of \autoref{fig:vis-grs-per-country-proportional}(top).}
	\label{fig:grs_all-per-country-comparison-proportional-and-absolute}
\end{figure*}

\begin{figure*}
	\centering
	\includegraphics[height=.53\columnwidth]{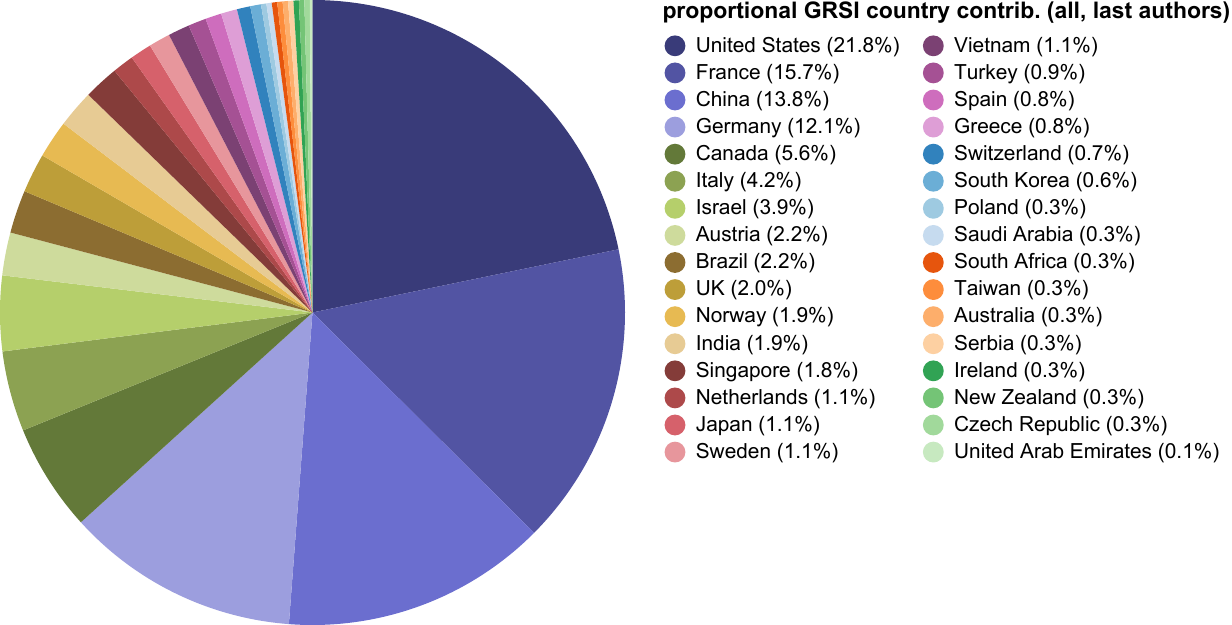}\hfill%
	\includegraphics[height=.53\columnwidth]{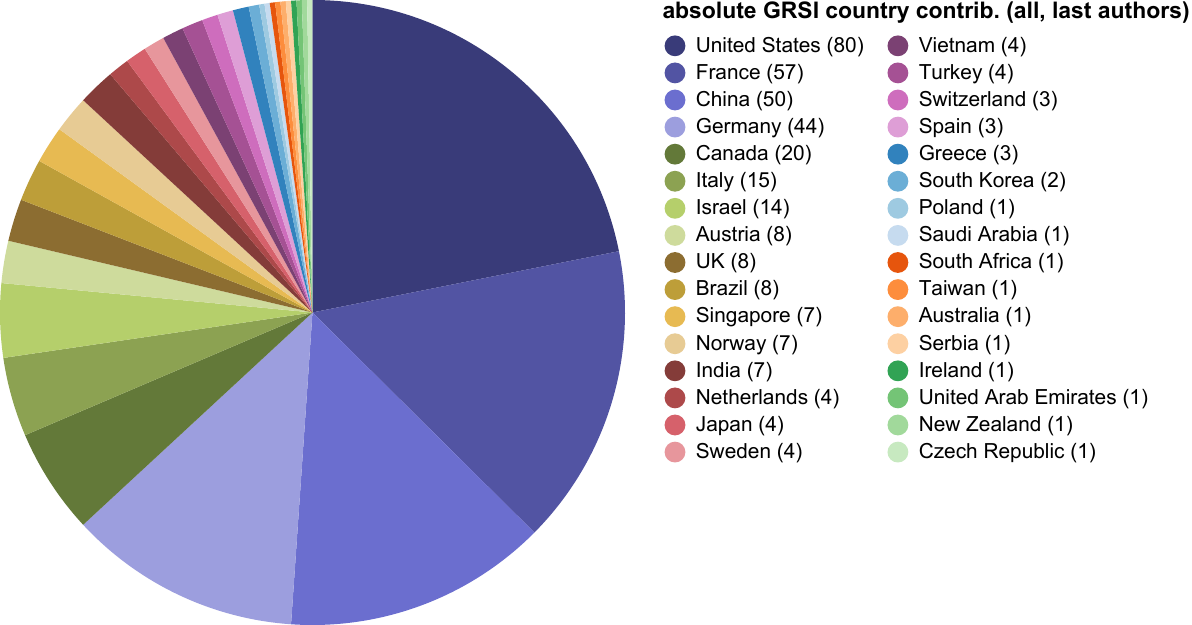}
	\caption{Proportional (left) and absolute (right) country contribution to GRS awards, looking only at last paper authors. The latter counts each paper with a co-author from a country for that country (each paper being counted once per country at most), while the former pro-rates the contributions among all co-authors of a papers and also potentially among their individual shared co-affiliations as described in \autoref{sec:visual_analysis}. The left image is a more detailed version of \autoref{fig:vis-grs-per-country-proportional-last-authors}(top).}
	\label{fig:grs_all_senior-per-country-comparison-proportional-and-absolute}
\end{figure*}

\begin{figure*}
	\centering
	\includegraphics[height=.53\columnwidth]{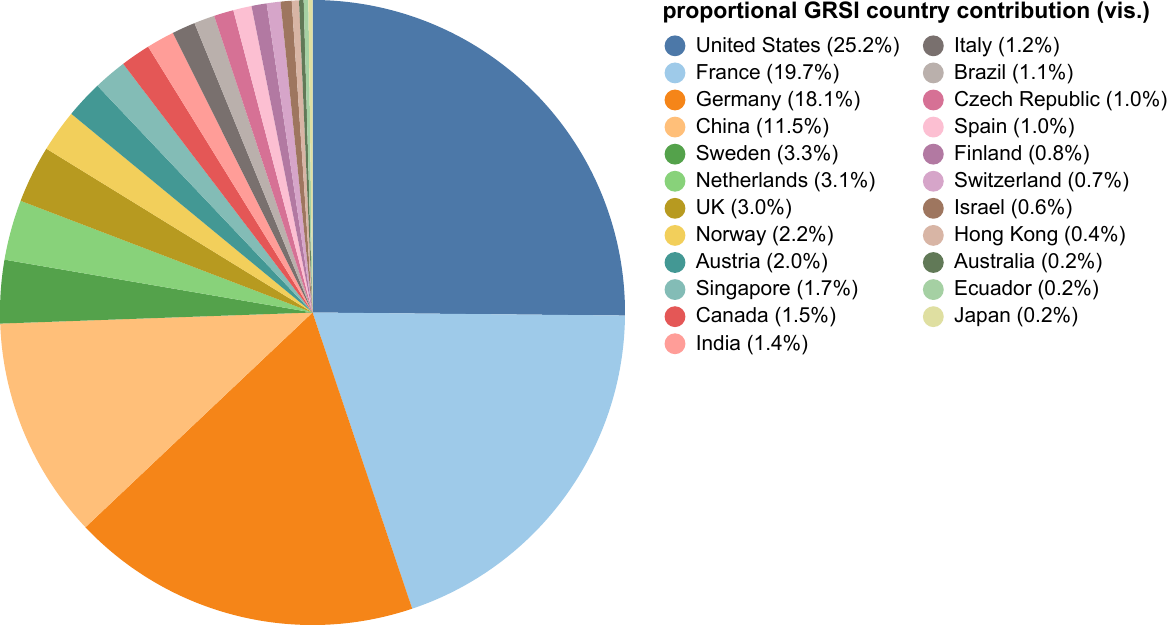}\hfill%
	\includegraphics[height=.53\columnwidth]{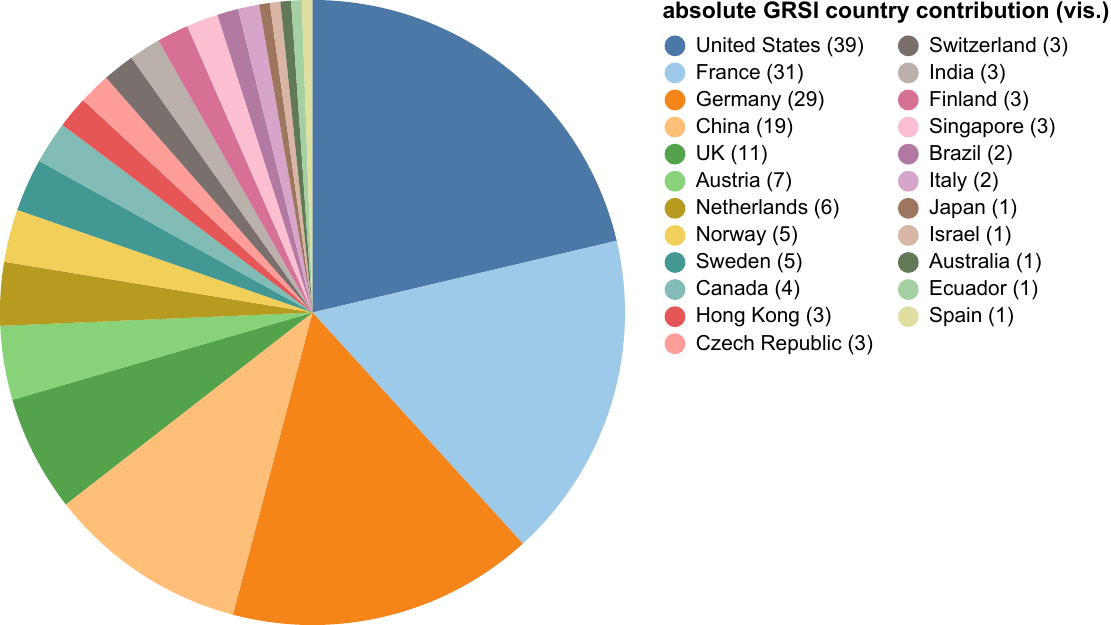}
	\caption{Proportional (left) and absolute (right) country contribution to GRS awards, only for visualization papers but looking at all paper authors. The latter counts each paper with a co-author from a country for that country (each paper being counted once per country at most), while the former pro-rates the contributions among all co-authors of a papers and also potentially among their individual shared co-affiliations as described in \autoref{sec:visual_analysis}. The left image is a more detailed version of \autoref{fig:vis-grs-per-country-proportional}(bottom).}
	\label{fig:grs_visualization-per-country-comparison-proportional-and-absolute}
\end{figure*}

\begin{figure*}
	\centering
	\includegraphics[height=.53\columnwidth]{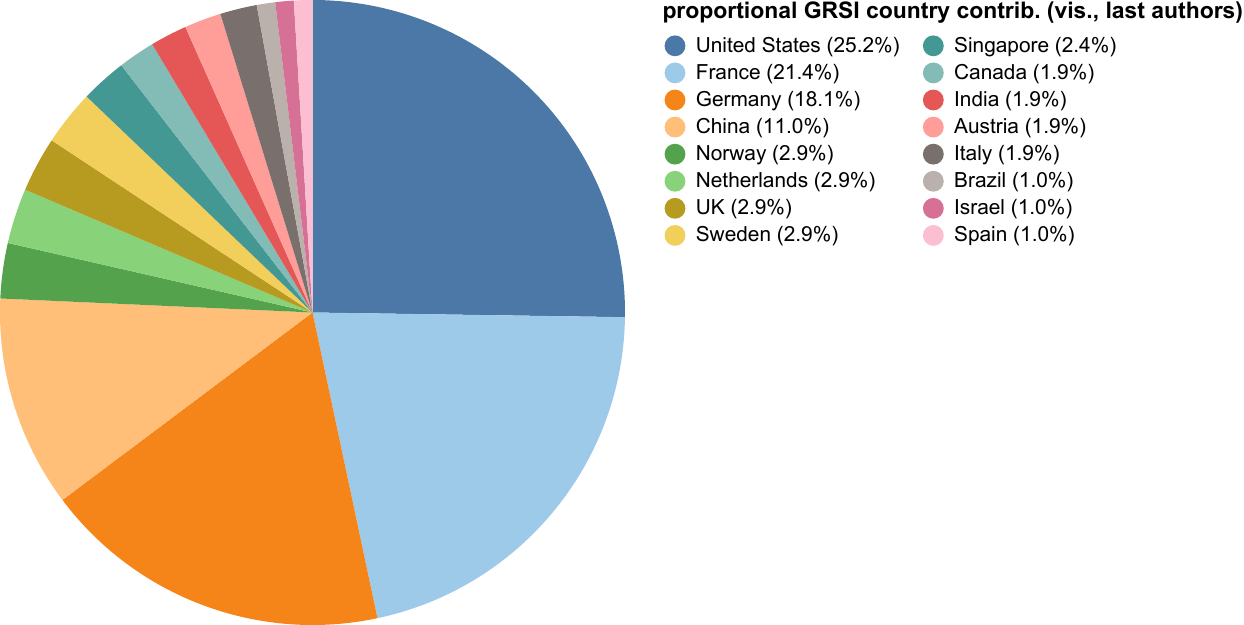}\hfill%
	\includegraphics[height=.53\columnwidth]{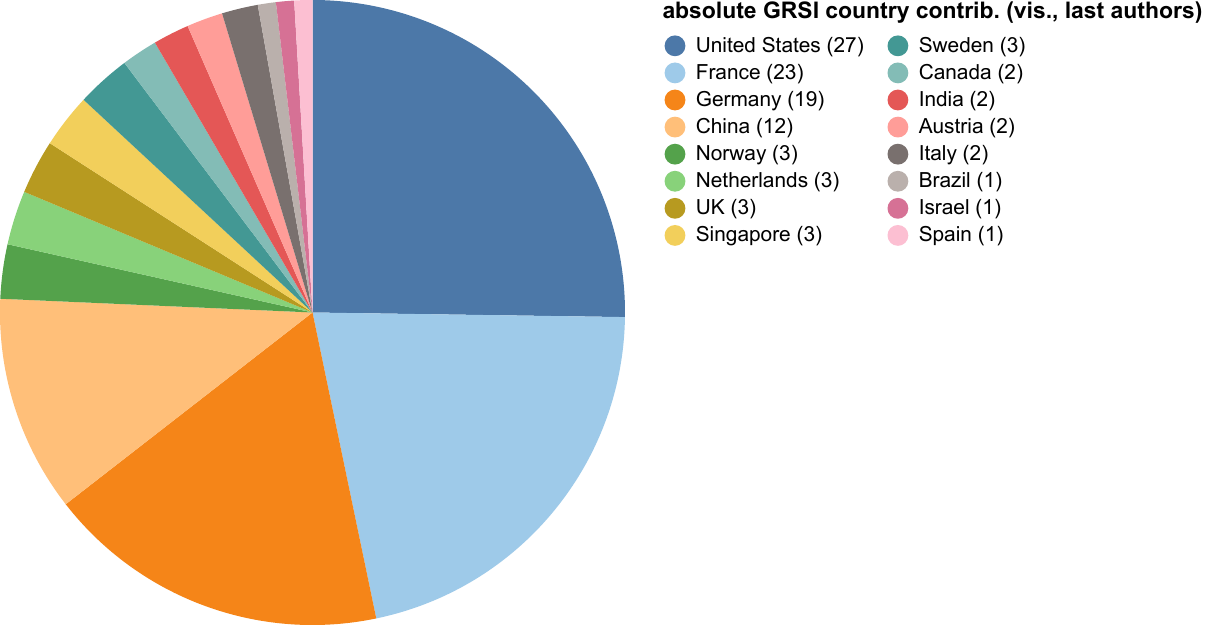}
	\caption{Proportional (left) and absolute (right) country contribution to GRS awards, only for visualization papers and only looking at last paper authors. The latter counts each paper with a co-author from a country for that country (each paper being counted once per country at most), while the former pro-rates the contributions among all co-authors of a papers and also potentially among their individual shared co-affiliations as described in \autoref{sec:visual_analysis}. The left image is a more detailed version of \autoref{fig:vis-grs-per-country-proportional-last-authors}(bottom).}
	\label{fig:grs_visualization_senior-per-country-comparison-proportional-and-absolute}
\end{figure*}

Below in \autoref{fig:vis-grs-presentations-by-year-linegraph}--\ref{fig:grs_visualization_senior-per-country-comparison-proportional-and-absolute} I include several additional views that show the data discussed in the main paper in different ways, such as normalized versions or graphs with more details than in the main paper. In the respective figure captions I explain the details for each figure and also link to the corresponding figures in the main paper.

\section*{Figure credits}
\label{sec:figure_credits_appx}

I as the author of this paper state that all figures in this appendix are my own as well as are and remain under my own personal copyright, with the permission to be used here. I also make them available under the \href{https://creativecommons.org/licenses/by/4.0/}{Creative Commons At\-tri\-bu\-tion 4.0 International (\ccLogo\,\ccAttribution\ \mbox{CC BY 4.0})} license and share them at \href{https://osf.io/\osfid/}{\texttt{osf.io/\osfid}}.

\end{document}